\documentstyle[amsmath,latexsym,graphicx]{jns}
\newcommand{\nc}{\newcommand} 
\nc{\diff}[2]{\frac{d #1}{d #2}}
\nc{\diffn}[3]{\frac{d^{#3} #1}{d {#2}^{#3}}} 
\nc{\pdiff}[2]{\frac{\partial #1}{\partial #2}} 
\nc{\pdiffn}[3]{\frac{\partial^{#3} #1}{\partial{#2}^{#3}}} 
\nc{\abs}[1] {\lvert #1 \rvert} 
\nc{\norm}[2] {{\lVert #1 \rVert}_{#2}} 
\nc{\threeline}[1] {\lvert \lvert \lvert #1 \rvert\rvert\rvert}
\nc{\cF}{{\cal F}} 
\nc{\cL}{{\cal L}} 
\nc{\cM}{{\cal M}}
\nc{\cN}{{\cal N}} 
\nc{\cE}{{\cal E}} 
\nc{\cZ}{{\cal Z}} 
\nc{\cT}{{\cal T}} 
\nc{\order}{{\cal O}}
\nc{\Eplus}{E_+} 
\nc{\Eminus}{E_-} 
\nc{\Epm}{E_\pm}
\renewcommand{\d}{\delta}
\nc{\w}{\omega} 
\nc{\eps}{\epsilon} 
\nc{\e}{\varepsilon} 
\nc{\g}{\gamma} 
\nc{\z}{\zeta}
\nc{\G}{\Gamma} 
\renewcommand{\k}{\kappa}
\nc{\pZ}{\partial_Z} 
\nc{\pT}{\partial_T} 
\nc{\pz}{\partial_z}
\nc{\pt}{\partial_t} 
\nc{\vu}{\Vec u} 
\nc{\vE}{\Vec {\cal E}} 
\nc{\vr}{\Vec r} 
\nc{\vrho}{\Vec \rho} 
\nc{\Reps}{R^{\e}}
\nc{\Vreps}{\Vec \Reps}
\nc{\half}{\frac{1}{2}}
\nc{\bphi}{\bar{\phi}} 
\nc{\efour}{{\Hat e}_4} 
\nc{\marginnote}[1] {\marginpar{\tiny #1}}

\newtheorem{cor}{Corollary} 
\newtheorem{thm}{Theorem} 
\newtheorem{prop}{Proposition}
\newtheorem{HN}{Hypothesis}
\DeclareMathOperator{\sech}{sech}
\DeclareMathOperator{\sign}{sign}
\DeclareMathOperator{\essup}{ess\, sup}
\DeclareMathOperator{\diag}{diag}

\title{ Nonlinear Propagation of Light in One Dimensional Periodic Structures}
\titlerunning{Nonlinear Light Propagation in 1D Periodic Media}

\author{
Roy H. Goodman\inst{1}\inst{2}\thanks{goodmanr@research.bell-labs.com}, 
Michael I. Weinstein\inst{1}\thanks{miw@research.bell-labs.com}, 
Philip J. Holmes\inst{2}\inst{3}\thanks{pholmes@rimbaud.princeton.edu} 
}
\authorrunning{R.H. Goodman, M.I. Weinstein, P.J. Holmes}
\institute{ Mathematical Sciences Research,  Fundamental Mathematics
Department, Bell Laboratories--Lucent Technologies, 600 Mountain Avenue, 
Murray Hill, NJ 07974
\and  Program in Applied and Computational Mathematics,
Princeton University, Princeton, NJ 08544
\and Department of Mechanical and Aerospace Engineering, Princeton University,
Princeton, NJ 08544}

\begin{document}
\maketitle
\setcounter{footnote}{0}

\begin{abstract}
We consider the nonlinear propagation of light in an optical fiber waveguide
as modeled by the anharmonic Maxwell-Lorentz equations (AMLE). The waveguide
is assumed to have an index of refraction which varies periodically along its
length.  The wavelength of light is selected to be in resonance with the
periodic structure (Bragg resonance).  The AMLE system considered incorporates
the effects of non-instantaneous response of the medium to the electromagnetic
field (chromatic or material dispersion), the periodic structure (photonic
band dispersion) and nonlinearity. We present a detailed discussion of the
role of these effects individually and in concert.  We derive the nonlinear
coupled mode equations (NLCME) which govern the envelope of the coupled
backward and forward components of the electromagnetic field.  We prove the
validity of the NLCME description and give explicit estimates for the
deviation of the approximation given by NLCME from the {\it exact} dynamics,
governed by AMLE.  NLCME is known to have gap soliton states.  A consequence
of our results is the existence of very long-lived {\it gap soliton} states of
AMLE. We present numerical simulations which validate as well as illustrate
the limits of the theory.  Finally, we verify that the assumptions of our
model apply to the parameter regimes explored in recent physical experiments
in which gap solitons were observed.
\end{abstract}
 
\section{Introduction}

There is a great deal of current interest in nonlinear optical phenomena in
periodic structures. This interest has been fueled by advances in fabrication
methods for periodic media and in their potential for use as components in
all-optical communication systems and computers.  The potential for
applications is due to the rich variety of phenomena which result from the
interactions of sufficiently intense (nonlinear) electro-magnetic fields with
the underlying (linear) dispersion characteristics of the periodic
structure~\cite{Y:93}. The reason one may envision the use of nonlinear
periodic structures in optical devices stems from the observation that one can
achieve very strong dispersion of a light pulse over very short distances by
arranging the wavelength of light and period of the medium to be appropriately
resonant.  At sufficiently high intensities, one then expects a balance
between nonlinear and dispersive effects over short distances, thus giving
rise to a rich class of phenomena in structures of small physical dimensions.

This paper is motivated by experiments and theory on nonlinear wave
propagation in one-dimensional periodic structures. Our goal is to validate
the nonlinear coupled mode equations (NLCME), a model commonly used to
describe this situation, and to clarify the roles played by the various
physical mechanisms.  The experiments involve the propagation of intense light
in an optical fiber waveguide whose core has a periodically varying index of
refraction along the length of the fiber, a {\it fiber grating}~\cite{K:99}.
Experimentalists have observed the formation of {\it gap solitons},
solitary-wave-like localized structures whose time-frequency parameters lie in
the photonic band-gap associated with the background periodic structure.
These are of potential interest for use in all optical storage devices, since
they can, in principle, travel at arbitrarily low speeds.  Theoretical work on
nonlinear propagation in periodic structures goes back to work of Winful
et. al.~\cite{WC:82,WMG:79}, and Chen and Mills~\cite{CM:87}.  Explicit gap
soliton solutions were derived in the context of a slowly varying envelope
theory by Christodoulides and Joseph~\cite{CJ:89} and in a more general form
by Aceves and Wabnitz~\cite{AW:89}. For surveys on aspects related to this
paper, see de Sterke and Sipe~\cite{DS:94}, Brown and Eggleton~\cite{BE:98}
and Kurizki et. al.~\cite{Kurizki:00}.  Experiments demonstrating the
existence of gap solitons have been performed by Eggleton, Slusher and
collaborators~\cite{E:97,E:99,E:96}, and by Broderick and his
collaborators~\cite{B_etal,M_etal}.  In two and three dimensions, Ak\"ozbek
and John have formally derived envelope equations and examined their solitary
waves numerically~\cite{AJ:98}.
 
In the remainder of this section we give brief overview of the underlying
physics and modeling assumptions. We also introduce the analytical and
numerical results developed in this article.

Electromagnetic wave propagation in a dielectric medium is described by
Maxwell's equations together with an appropriate constitutive relation
describing how electromagnetic waves interact with matter. An optical fiber
has a high index {\it core} and a slightly lower index {\it cladding}. This
index configuration confines rays to the core (total internal reflection) or,
from the wave perspective, the index profile provides a potential well with a
ground state (core mode) having most of its energy confined to the core.  In
the regimes which interest us here, to a very good approximation the energy
distribution has a fixed transverse structure given by the core mode and one
may think of the transverse core mode amplitude as varying with time $t$ and
distance along the waveguide, $z$. In addition to this geometric constraint,
we incorporate the following effects:

(i) {\it Non-instantaneous response of the medium to the field:} The
polarization, $P$, is related to the electric field, $E$, via an anharmonic
Lorentz oscillator model.

\begin{figure}
\begin{center}
\includegraphics[width=12.2cm]{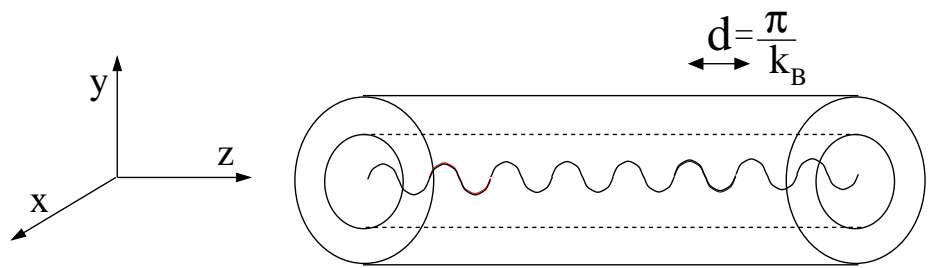}
\caption{A schematic drawing of an optical fiber with periodic refractive
index variation} 
\end{center}
\label{fig:physical_schematic}
\end{figure}

(ii) {\it Periodicity of the medium:} Spatial periodicity of the medium 
is built in by allowing the coefficients of the anharmonic Lorentz oscillator
to vary periodically in space. A schematic of the physical system is shown in
Figure~\ref{fig:physical_schematic}. 

(iii) {\it Nonlinear effects at appropriate intensities:} The implied relation
between $P$ and $E$ is such that regions of higher intensity $|E|^2$ have
higher refractive index. This is a so-called focusing (Kerr) nonlinearity.
The localized region of higher intensity effectively creates an attractive
potential well.  

The effects of non-instantaneous response and spatial periodicity each give
rise to {\it dispersion}, the property that waves of different wavelengths
travel at different speeds.  The type of dispersion due to (i) is called {\it
chromatic or material dispersion} and that arising due to effect (ii) is
called {\it photonic band dispersion}. This results from interference effects
arising from reflection and transmission in the periodic structure.

A model incorporating the above geometric constraints and physical effects is
a variant of the Anharmonic Maxwell-Lorentz system~\cite{B:92,O:71,JMR:96},
which incorporates the spatial periodicity. This system is displayed
in~\eqref{eq:AMLE} and we shall refer to it below as AMLE.\footnote{In the
nonlinear optics literature, the relation between polarization $P$ and
electric field $E$ is often taken to have the form: $P = \int_{-\infty}^t
\chi(t-\tau) E(\tau) d\tau + \dots$. The class of models we have chosen gives
the same envelope equation, NLCME, in the scaling regime considered but has
the added feature that it conserves energy.  Energy estimates are central to
our proof of the validity of NLCME: Theorem~1.} 

\begin{figure}\begin{center}
\includegraphics[width=4in]{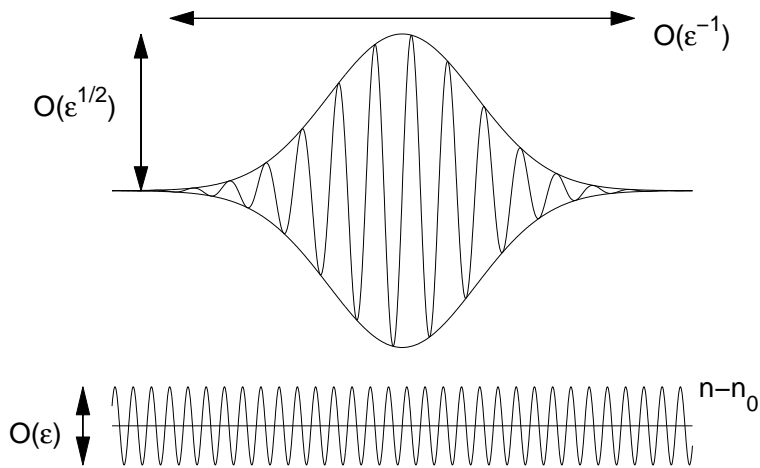}
\caption{A schematic of a wave packet under the SVEA, with envelope with width
$\e^{-1}$, amplitude $\sqrt{\e}$, carrier wavelength $2d$ 
and a plot of the index of
refraction variation with variation of size $\e$ and period $d$.}
\label{fig:schematic}
\end{center}\end{figure}

While in a bare (homogeneous, non-grated) optical fiber, light injected at one
end of the waveguide will propagate with little back-scatter, significant
back-scattering will occur in the presence of a periodic refractive index.
This effect is most pronounced when the wave-length of light is roughly twice
the grating period, $2d$, the case of {\it Bragg resonance}.  In this case
there is strong coupling between backward and forward waves. We will assume
that the variation of the index of refraction about its mean is small and is
denoted by a parameter $\e$.  In terms of this parameter we consider the
following scaling regime:
\begin{quote}
 $\bullet$\  amplitude of the field  $\sim\ \order(\sqrt{\e})$, and 
\end{quote}
\begin{quote}
$\bullet$\ initial spectral support of the pulse is concentrated in
a wave number range of width $\order(\e)$ about $\pm k_B\equiv \pm \pi/d$.  
\end{quote}
Therefore, the spatial structure of the fields $E$ and $P$
may be viewed as functions of the form $\sqrt{\e}A(\e x) e^{ik_Bx}$, where
$A(y)$ is a localized function of $y$.  We shall refer to this as the {\it
slowly varying envelope approximation} (SVEA).  A schematic of this scaling
ansatz is shown in Figure~\ref{fig:schematic}.

Under these assumptions, {\em nonlinear coupled mode equations} (NLCME, see
equations~\eqref{eq:cme}) can be derived which govern the forward $E_+$ and
backward $E_-$ propagating electric field wave envelopes on time scales of
order $\e^{-1}$.  Thus, in this regime, the fine scale grating oscillations
are effectively averaged and the original mathematical description in terms of
a nonlinear partial differential equation with spatially periodic coefficients
is replaced by a constant coefficient dispersive nonlinear partial
differential equation.

The main results of this paper are as follows:

\noindent{\bf Characterization of Phenomena:} The formation of long-lived
coherent structures (gap solitons) is the result of a balance between the
effects of dispersion and nonlinearity.\footnote{In contrast to the case of
bare fiber, for which this balance is achieved over lengths of optical fiber
on the order of tens of kilometers, for the periodic structures and
intensities used in the above cited experiments, this balance occurs on a
length scale of centimeters; see Appendix B.}  The energy in a wave packet,
with frequency content localized about the Bragg resonant frequency, resolves
into backward and forward propagating waves. If the field amplitude is
appropriately large relative to the the amplitude of periodic variations in
the medium, then wave energy does not disperse and is localized in
space. Nonlinearity generates ever higher harmonics which is manifested in
wave steepening and apparent carrier shock formation; see
Sects.~\ref{sec:aperiodic_nl} and~\ref{sec:periodic_nl}.  However, in the
presence of material dispersion due to finite time response, a stable balance
between dispersion and nonlinearity is achieved; no shocks form and one has
long-lived stable gap solitons (see Sect.~\ref{sec:nl_finite_time},
Theorem~\ref{thm:main}, and the corollary of
Sect.~\ref{sec:periodic_nl_finite}).  We also verify
(Appendix~\ref{sec:nondim}) that parameter ranges corresponding to experiments
are described by our model and theorem.

\noindent {\bf Analytical---Theorem~\ref{thm:main}}: We prove that solutions
to the initial value problem for AMLE with finite energy and nearly
monochromatic initial conditions, as above, give rise to solutions which are
well-approximated, on appropriate time scales ($\order(\e^{-1})$) by a
superposition of amplitude-modulated backward and forward propagating plane
waves. The backward and forward amplitudes satisfy NLCME. Further, we estimate
the deviation of the NLCME approximation to the AMLE solution.  An important
class of solutions of NLCME are so-called {\it gap solitons}.\footnote{Here,
we adopt a common usage of the term soliton as referring to a nonlinear bound
state solution or solitary wave. The term originally and still often refers
more specifically to nonlinear bound states arising in completely integrable
systems.}  These are spatially localized nonlinear bound states~\cite{AW:89}
which have been observed experimentally. Our results imply the existence of
gap soliton wave packets for AMLE on time scales $\order(\e^{-1})$, see
Sect.~\ref{sec:validity}. Our method follows previous rigorous studies of the
validity of solutions to envelope equations in approximating oscillatory,
nearly monochromatic solutions to evolution PDE's in one space dimension; see,
for example, Kirrman--Schneider--Mielke~\cite{KSM:92} and
Pierce--Wayne~\cite{PW:95}.  A related method was presented in the context of
dissipative equations is given by van Harten~\cite{vH:91}.  Important
extensions of such methods have been developed in Donnat--Rauch~\cite{DR:97},
Joly--Metivier--Rauch~\cite{JMR:98}, Lannes~\cite{L:98}, and
Schochet~\cite{S:94b, S:94a} where multidimensional and multiphase problems
are treated. These paperes do not treat the resonant interactions of
multi-phase waves with periodic media, although the formulation
of~\cite{DR:97}, e.g., can (easily) be extended to deal with the present
one-space-dimensional problem. Here, however, we provide a self-contained,
elementary treatment of the one-dimensional problem.  We also wish to note a
recent general paper of Babin and Figotin~\cite{BF:00} on wave interactions in
periodic media.

\noindent{\bf Numerical Simulations:} We numerically simulate AMLE and NLCME
and systematically compare their computer generated solutions with a view
toward checking the error estimates of Theorem~\ref{thm:main}.  Initial data
appropriate for Theorem~\ref{thm:main} consists of a slow modulation of a
highly oscillatory wave. In the main numerical example presented in
Sect.~\ref{sec:numerics}, we have simulated the AMLE evolution with gap
soliton wave packet initial data.  We take data with an envelope whose full
width at half maximum (FWHM) measured in wavelengths of light, is about 60,
see Figure~\ref{fig:efield}.  In one run, the simulation of AMLE took about
$2\half$ hours, while the corresponding simulation of NLCME took only a few
minutes on a 500 mHz Pentium III computer running Linux.  The advantage is
even larger when wavepackets with more oscillations are investigated.  For
parameter regimes of physical interest, it is probably infeasible to simulate
the full AMLE, while it is quite simple to simulate the NLCME. In the physical
experiments~\cite{E:97}, pulses on the the order of 30 ps FWHM are observed,
$\order(10^4)$ wavelengths.

In our numerical simulations we identify three time regimes. The first is the
time scale on which the coherent structure evolves as a gap soliton plus
fluctuations satisfying the estimates of Theorem~\ref{thm:main}.  The second
is a longer time scale on which the wave envelope predicted by NLCME gives a
accurate prediction of where the field energy is, but due to phase drift, the
norm estimates of the error in Theorem~\ref{thm:main} fail to hold. The third
is a regime on which the wave envelope begins to steepen asymmetrically and
radiate energy, leading to a decay of the gap soliton. A description of this
process would require the inclusion of higher order nonlinear wave-steepening
and dispersive corrections to NLCME.  
 
The structure of this paper is as follows:

In \S\ref{sec:AMLE} we introduce the anharmonic Maxwell-Lorentz model (AMLE)
and inn \S\ref{sec:NLCME} we describe and display the nonlinear coupled mode
equations, discuss their mathematical structure, and state our main theorem
(Theorem~\ref{thm:main}) relating solutions of AMLE to those of NLCME. In
\S\ref{sec:phenomena} we discuss the effect of the nonlinearity, periodic
structure, and material dispersion and describe the physical effects of
including, excluding and variously combining these mathematical features of
the system. In \S\ref{sec:derivation} we present a derivation of NLCME from
AMLE using the method of multiple scales and in \S\ref{sec:ivp} we discuss
existence and uniqueness results for AMLE and NLCME, some of which are needed
in \S\ref{sec:validity} where we prove Theorem~\ref{thm:main}. In
\S\ref{sec:numerics} we report on numerical simulations and careful systematic
comparison of solutions to AMLE and NLCME.  In \S\ref{sec:summary} we
present a short summary followed by a discussion of issues meriting further
investigation.  The appendices contain a discussion of nondimensionalization
and physical parameter magnitudes, and details of dimensionless values used in
the numerical simulations.

\medskip

\noindent{\bf Notation and Conventions}\\
Throughout this paper we make use of following notation:

The symbols $C, C_j$ are used to represent generic constants whose
dependence on parameters is specified when of concern.  

For a vector-valued function $\Vec f(z)$, the $L^p$-norm is given by
\begin{equation}
\norm{\Vec f}{p} = \bigl( \int \sum_j \abs{f_j(z)}^p dz \bigr)^{1/p}\enspace.
\end{equation}
Here and throughout, spatial ($z$) integrals are taken over $-\infty<z<\infty$.
The space $L^p$ is then the space of all functions $\Vec f$ such that 
$\norm{\Vec f}{p}$ is finite.

The $L^\infty$ norm is given by
\begin{equation}
\norm{\Vec f}{\infty} = \max_j \left(\essup |f_j(z)|\right)
\end{equation}
with the space $L^\infty$ thus defined as the set of all (essentially) bounded
functions. 

The $H^s$ norms may be defined as
\begin{equation}
\norm{\Vec f}{H^s}^2 = \sum_{k=1}^{s} \Bigl\lVert\diffn{}{x}{k}\Vec f
\Bigr\rVert_2^2\enspace.
\end{equation}
The space $H^s$ is the space of all functions $\Vec f$ such that 
$\norm{\Vec f}{H^s}$ is finite, that is, the space of functions such that the
function and its first $s$ derivatives are square-integrable.

Finally, for a given Banach space $X$, with norm $||\cdot ||_X$, we define
$$
C([0,T); X)
$$
to be the set of functions $f:t\mapsto f(t) \in X$ 
which  are continuous for $t\in [0,T)$ with values in $X$.

\section{The AMLE and NLCME Equations}
\subsection{AMLE, nondimensionalization and parameter regimes}
\label{sec:AMLE}
In this subsection we introduce AMLE with physical parameters and then 
introduce its nondimensional form. We then discuss parameter regimes
associated with the above mentioned experiments. 

We take as our basic model a one-dimensional electromagnetic system satisfying
Maxwell's equation, with the polarization governed by an anharmonic Lorentz
oscillator model,%
\footnote{ Our results and analysis apply to the generalization of this model
where we take $P$ to be a weighted sum of $N$ polarizations, $P_i$,
corresponding to different  molecular excitation modes of the material:
\begin{equation}
P= \sum_{i=1}^{N} P_i \enspace ; \qquad
\tilde\w_i^{-2}\pt^2 P_i + 
\left(1-2 \Delta n_i \cos(2\tilde k_Bz)\right)P_i -\tilde\phi_i P_i^3 
= \eps_0 \chi_i^{(1)}E \enspace. \nonumber
\end{equation}
This model can be viewed as a nonlinear generalization of the Sellmeier
model~\eqref{eq:sellmeier}; see~\cite{A:95}.  With this particular modeling of
the polarization, AMLE has the important property of being an energy
conserving system. This structure gives rise to {\it energy estimates} which
are central to our proofs of well-posedness of AMLE and of the validity of
NLCME as an approximating envelope equation; see Sects.~\ref{sec:ivp}
and~\ref{sec:validity}.}  henceforth referred to as the anharmonic
Maxwell-Lorentz equations (AMLE)~\cite{B:92,JMR:96,O:71}
\begin{subequations}
\begin{align}
&\mu_0 \pt D = \pz B;\;  \pt B  = \pz E \enspace;
\\
&D \equiv \eps_0 E+P\enspace ;\\
&\tilde\w_0^{-2}\pt^2 P + 
\left(1-2 \Delta n \cos(2\tilde k_Bz)\right)P -\tilde\phi P^3 
= \eps_0 \chi^{(1)}E \enspace.
\end{align}
\end{subequations}
Here, $E$ is the electric field, $B$ is the magnetic field, $P$ is the
polarization, and $D$ is the electric displacement. $\epsilon_0$ and $\mu_0$
denote, respectively the permitivity and permeability of free space, and $\chi^{(1)}$ is the linear polarizability of the medium. Recall
that $\epsilon_0\mu_0=c^{-2}$, where $c$ is the vacuum speed of light. $\Delta
n$ measures the strength of the grating. We shall also write:
\begin{equation}\Delta n = \e\nu \enspace,\end{equation} 
where $\e$ measures the size of the index modulation and $\nu$ is of order 
one and is introduced in order to make explicit how the spatially periodic
structure rears its head in the envelope approximation, NLCME, to be derived
below. 

The spatial period of the medium is $d$ and is expressed in terms of 
$$\tilde k_B=\frac{\pi}{d}\enspace,$$
Since we are interested in the propagation of light with wavelength equal to
the Bragg wavelength, we set
$$\lambda= 2d \enspace \enspace,$$ 
where $d$ denotes the period of the grating.

In Appendix~\ref{sec:nondim}, we eliminate the magnetic field $B$ from this
system, and nondimensionalize these equations. There, nondimensional dependent
variables are primed, but here we drop primes for simplicity of notation.
From~\eqref{eq:wave_eq_nondim}, \eqref{eq:Ddef}, and~\eqref{eq:P_eqn_nondim},
we obtain:
\begin{subequations}
\label{eq:AMLE}
\begin{align}
&\pt^2 D = \pz^2 E \enspace ;
\label{eq:AMLE1} \\
&D \equiv E+P \enspace ;\label{eq:D_eqn}\\
&\w_0^{-2}\pt^2 P + (1-2 \e\nu \cos(2k_Bz))P -\phi P^3 
\label{eq:AMLE2} = (n^2-1)E\enspace.
\end{align}
\end{subequations}
$\w_0$ is a dimensionless frequency and $\phi$, a dimensionless measure of
the degree of the nonlinearity.  The limit of instantaneous polarization is
achieved by taking the parameter $\w_0 \to \infty$. This gives the relation
$$P = P(E) = (1+ 2\e\nu\cos(2k_Bz))E + \chi^{(3)} E^3 + \dots$$ with nonlinear polarizability $\chi^{(3)} $
and, in this case, equation~\eqref{eq:AMLE} 
reduces to the scalar nonlinear wave equation.
\begin{equation}
\pt^2( n^2E + 2\e\nu\cos(2k_Bz) E + \chi^{(3)}E^3) = \pz^2 E
\enspace.
\label{eq:scalarnlwe}
\end{equation}

\subsection{NLCME and Main Results}
\label{sec:NLCME}
The nonlinear coupled mode equations are introduced by considering slow
modulations to solutions of the anharmonic oscillator model in which the
photonic structure and nonlinearity are ignored.  When $\e=0$ and $\phi=0$
system~\eqref{eq:AMLE} supports plane wave solutions of the form $E= \Epm
e^{i(\pm kz -\w t)}$, where $k=k(\w)$ is the dispersion relation (see
Sect.~\ref{sec:phenomena}) and $E_\pm$ are constants.  A similar statement
holds for $P$. In the scaling regime described in the introduction, in which
nonlinear effects and spatial periodicity are allowed, and where the carrier
wave has wavenumber $k_B$ and frequency $\w_B = \w(k_B)$, we seek
coupled and {\it slowly modulated} backward and forward plane wave solutions
of the form
\begin{equation}
\label{eq:lead}
\binom{E_{NLCME}^\e}{P_{NLCME}^\e} \sim  
 \sqrt{\e}\left( \Eplus(Z,T) e^{i(k_Bz-\w_B t)} 
   + \Eminus(Z,T) e^{-i(k_B z+\w_B t)} + c.c.  \right)\binom{1}{\gamma_B}
 \enspace, 
\end{equation} 
where $c.c.$ denotes the complex conjugate of the previous expression and
$\gamma_B = \gamma(\w_B)$ is a constant.
Here, $Z$ and $T$ are ``slow variables:''
\begin{equation} 
Z= \e z, \qquad T= \e t\enspace ,
\end{equation}
and $\Eplus$ and $\Eminus$ satisfy equations of the form%
\footnote{Beginning with a three dimensional Maxwell-Lorentz model in fiber
geometry, it is possible to derive similar nonlinear coupled mode equations,
with one difference being that the term $\G(\abs{E_\pm}^2+
2\abs{E_\mp}^2)E_\pm$ is replaced by one of the form
$(\G_s\abs{E_\pm}^2+2\G_\times\abs{E_\mp}^2)E_\pm$, where $\G_s$ and
$\G_\times$ are the nonlinear self-phase modulation and cross-phase modulation
coefficients, and depend on certain integrals of the transverse modes of the
waveguide.~\cite{AW:89,NM:92} Another difference one finds is that the
transverse potential, defined by the refractive index profile, induces {\it
waveguide or modal dispersion}. Thus, a more complete description of the
physics leads to corrections to the free space dispersion relation due to
material dispersion, photonic band dispersion and modal dispersion, as well.
The multidimensional analyses of~\cite{DR:97, JMR:98, L:98, S:94a, S:94b},
which assume ``almost'' plane wave solutions, do not appear to generalize
(easily) to the waveguide problem.}
\begin{subequations}
\label{eq:cme}
\begin{align}
i \left( \pT\Eplus +  v_g \pZ\Eplus \right) + \k \Eminus +
\G(\abs{\Eplus}^2 + 2\abs{\Eminus}^2)\Eplus &= 0 
\enspace ;\label{eq:cme1}\\ 
i \left( \pT\Eminus -  v_g \pZ\Eminus \right)  + \k \Eplus +
\G(\abs{\Eminus}^2 + 2\abs{\Eplus}^2)\Eminus &= 0
\enspace . \label{eq:cme2}
\end{align}
\end{subequations}
Here, $\k$ is a coupling parameter (proportional to $\nu$ induced by the
grating), $v_g$ is the group velocity of the linear dispersive wave at
frequency $\w_B$, and $\G$ is the nonlinear coupling parameter (proportional
to $\phi$). The explicit expressions for these coefficients are displayed in
Sect.~\ref{sec:derivation}, during the derivation of~\eqref{eq:cme}; 
see~\eqref{eq:kgamma}. 

The expression in~\eqref{eq:lead} for $E^\e$ and $P^\e$ is a formal
approximate solution to AMLE satisfying the ``nearly monochromatic'' initial
condition:
\begin{equation}
\binom{E(z,t=0)}{P(z,t=0)} = \sqrt{\e} \left( E_{0+}(\e z){\Vec v_+} e^{ikz}
 + E_{0-}(\e z){\Vec v_-}e^{-ikz} + c.c. \right) + \order(\e)
\label{eq:amldata}
\end{equation}
where ${\Vec v}_\pm$ are constant two-component vectors.

We prove the following result in Sect.~\ref{sec:validity}:
\begin{thm}
Consider AMLE with a general nonlinearity satisfying Hypothesis~\ref{hyp:N2}
of Sect.~\ref{sec:validity}.  There exists $\e_0>0$ such that for any $T_0>0$
and any $0<\e\le\e_0$, the solution $\binom{E_{AMLE}^\e}{P_{AMLE}^\e}$ of AMLE
with data~\eqref{eq:amldata} belonging to $H^3$ is well approximated by a
solution of NLCME in the sense that for all $t\in [0,T_0/\e]$ the following 
estimate holds:  
\begin{equation}
\Biggl \lVert
{\binom{E_{AMLE}^\e}{P_{AMLE}^\e}-\binom{E_{NLCME}^\e}{P_{NLCME}^\e}}
\Biggr \rVert_{H^1} 
 \le C(T_0;\w_0, \nu, n) \e \enspace. 
\end{equation} \label{thm:main}
\end{thm}
\medskip

We note that due to Sobolev's inequality, $|f(x)|\le C||f||_{H^1}$, a small
error in the $H^1$ norm ensures a small pointwise error, so that the above
statement gives uniform bounds on the error.

\section{ AMLE, NLCME and Physical Phenomena}
\label{sec:phenomena}
The physical phenomena modeled by AMLE and NLCME result from competition
among: (i) nonlinear effects, (ii) dispersion due to finite time response of
the medium to the field and (iii) dispersion due to reflection and
transmission in a spatially periodic medium. This section is divided into
subsections in which we study, by considering various choices of $\w_0$, $\e$
and $\phi$ in~\eqref{eq:AMLE}, the action of these effects (terms in the
equations) individually and in concert.

\subsection{Linear spatially homogeneous structure with instantaneous
response:}   
\label{sec:linear_homogeneous}
In this case, $\e=\phi=0$ and $\w_0=\infty$. Therefore $P = (n^2-1)E$ and
the evolution is described by the classical wave equation:
\begin{equation}
n^2 \pt^2 E = \pz^2 E \enspace ,
\label{eq:we}
\end{equation}
whose solutions are of the form
\begin{equation}
E(z,t) = e_+(z-t/n) + e_-(z+t/n) \enspace,
\label{eq:leftright}
\end{equation}
corresponding to a a superposition of left and right moving waves which
propagate without distortion. 

Alternatively, we can first seek elementary plane wave solutions
$E(z,t;k)=e^{-i\w t +i kx}.$
We then find that $\w$ and $k$ are related by the simple {\it dispersion 
relation} $\w(k) = \pm \frac{k}{n}$. Since the phase velocity, 
$\w(k)/k$,  is 
independent of $k$, all wavelengths travel at the same speed and we refer
to~\eqref{eq:we} as {\it nondispersive}. Standard Fourier superposition
of these plane waves yields the general solution~\eqref{eq:leftright}.

\subsection{ Linear and homogeneous medium with finite time response}

In this case we have
\begin{subequations}
\begin{align}
\pt^2(E+P) &= \pz^2 E\enspace;\\
\w_0^{-2}\pt^2 P + P &= (n^2-1)E \enspace.
\end{align}
\end{subequations}
We may still find plane wave solutions
\begin{equation}
\binom{E_0}{P_0} = \binom{1}{\g} e^{i(kz-\w t)}\enspace , \label{eq:E0P0}
\end{equation}
where $k$ and $\w$ are related by the {\it dispersion relation}
\begin{equation}
k^2 =\w^2 \frac{n^2- \left(\dfrac{\w}{\w_0}\right)^2}
{ 1-\left(\dfrac{\w}{\w_0}\right)^2} \enspace , \label{eq:disp_rel}
\end{equation}
and
\begin{equation} \g = \frac{n^2-1}{ 1-\left(\dfrac{\w}{\w_0}\right)^2}\enspace.
\label{eq:gamma}\end{equation}
In the relevant parameter regimes $\w_0^2 > \w^2$ and $n^2-1>0$. So, $\g$ is
positive, corresponding to polarization in phase with the electric field.
Note that in the $\w_0 \to \infty$ limit, we recover the wave equation
dispersion relation $k = \pm \w n$.

A general solution may be constructed by superposition of these plane waves
using the Fourier Transform. Using the method of stationary phase~\cite{W:74},
one can show that for initial data whose Fourier transform decays sufficiently
rapidly, the amplitude of the solution decays  as $t^{-\half}$.  

\subsection{Linear periodic structure, instantaneous response}

In this case, we take $\phi=0$ and $\e\ne0$. We consider the case of
instantaneous response, $\w_0=\infty$, though the methods apply to finite
$\w_0$ as well. In this case we have the scalar one-dimensional wave
equation with spatially periodic wave speed: 
\begin{equation}
\left( n^2 + 2\e(n^2-1)\nu\cos(2k_Bz)\right)\pt^2 E = \pz^2 E \enspace.
\label{eq:perwave}
\end{equation}
In analogy with the scalar and spatially homogeneous wave equation, we seek
solutions of the form: $E(z,t) = e^{-i\w t} \varphi(z)$. This yields the
Mathieu equation:
\begin{equation}
-\pz^2\varphi(z) =
 \w^2\left( n^2 + 2\e(n^2-1)\nu\cos(2k_Bz)\right)\varphi(z)\enspace.
\label{eq:hill}
\end{equation}
We now seek solutions of~\eqref{eq:hill} of the form 
\begin{align}
\varphi(z)&=e^{iK z}\psi(z;K)\enspace , \qquad K\in [0,2\pi)\\
\psi(z+d;K) &= \psi(z;K) \enspace ,
\end{align}
where $\psi$ has the same periodicity as the medium.
Therefore, $\psi(z;K)$ satisfies the boundary value problem:
\begin{align} \label{eq:bvp}
-\left(\pz + iK\right)^2\psi(z;K) &=
  \w^2\left( n^2 + 2\e(n^2-1)\nu\cos(2k_Bz)\right) \psi(z;K) \enspace ;\\
 \psi(z+d;K) &= \psi(z;K), \ \ d\equiv \frac{\pi}{k_B} \enspace.
\end{align}
For each $K$, there is a discrete set of eigen-solutions 
$\{ \varphi_m(z;K): m=1,2,\dots\}$
with corresponding eigenvalues $\{ \w_m(K)^2: m=1,2,\dots\}$. 
As $K$ varies over the interval
$[0,2\pi)$, the functions $\w_m^2(K)$ sweep out spectral (photonic) bands. 
These bands are separated by spectral (photonic band) gaps. The solutions
\begin{equation}
E_m(z,t;K) = e^{-i\w_m(K)t + iKz} \psi_m(z;K),\ K\in [0,2\pi),\
  m=1,2,\dots\enspace.
\end{equation}
are generalizations of plane wave solutions of the previous subsections.  In
contrast to the homogeneous medium case, where the allowable set of
frequencies varies over the entire real line, in the periodic case the
allowable set of frequencies varies over selected bands.  The band dispersion
relations, $\{ \w_m(K): m=1,2,..\}$, play the role of the dispersion relation,
$\w(K)$, for the homogeneous medium (constant coefficient partial differential
equation). Since the phase velocities $\w_m(K)/K$ are not independent of $K$,
we see that wave propagation in periodic media is dispersive. A generalization
of Fourier superposition holds, enabling one to construct the general solution
to the initial value problem for~\eqref{eq:perwave}. A careful stationary
phase analysis of this superposition formula can be made, yielding results on
the spreading and temporal decay of solutions~\cite{Ko:97}.

Thus, Floquet-Bloch theory gives a complete characterization of wave
propagation in a linear periodic medium. However, the key to understanding the
detailed properties of this propagation is a detailed knowledge of the band
dispersion functions, $\w_m(K)$. This is a difficult problem, in general. In
the case when the periodicity is given by a small oscillation about its mean
($\e$ small), {\it coupled mode theory}~\cite{K:69} can be used to
approximate the Floquet-Bloch spectral theory. This provides a satisfactory
description of the wave propagation for large, but finite, times. We
illustrate this for equation~\eqref{eq:perwave}.  The idea is that for $\e$
small, the solutions $E_m(z,t;K)$ should be well-approximated by plane waves
of the unperturbed $\e=0$ problem.  Thus, we seek solutions
of~\eqref{eq:perwave} in the form:
\begin{equation}
\label{eq:cm-ansatz}
E = \left( \Eplus(\e z,\e t) e^{ik_B(z-t/n)}
      +\Eminus(\e z,\e t) e^{-ik_B(z+t/n)} + c.c.  \right) 
+\order(\e)\enspace ,
\end{equation}
and derive equations for the slowly varying functions $E_+(Z,T), E_-(Z,T)$,
ensuring that~\eqref{eq:cm-ansatz} is a good approximation of a solution for 
times, $t$, of order $\e^{-1}$. This approximation and error estimates are
derived systematically in Sects.~\ref{sec:derivation} and~\ref{sec:validity}
in the nonlinear context of AMLE. In this 
linear setting the equations reduce to the {\it linear coupled mode equations}
 ({\it cf.} equations~\eqref{eq:cme}):
\begin{subequations}
\label{eq:linearcme}
\begin{align}
i \left( \pT\Eplus +  v_g \pZ\Eplus\right) + \k \Eminus 
 &= 0  \enspace ; \label{eq:linearcme1}\\
i \left( \pT\Eminus - v_g \pZ\Eminus \right) + \k \Eplus 
 &= 0 \enspace, \label{eq:linearcme2}
\end{align}
\end{subequations}
where $v_g\equiv\w'=n^{-1}$ is the group velocity (which happens to agree here
with the phase velocity, $\w(k)/k$), and $\k=\frac{k(n^2-1)\nu}{2n}$.

\begin{figure}\begin{center}
\includegraphics[width=3in]{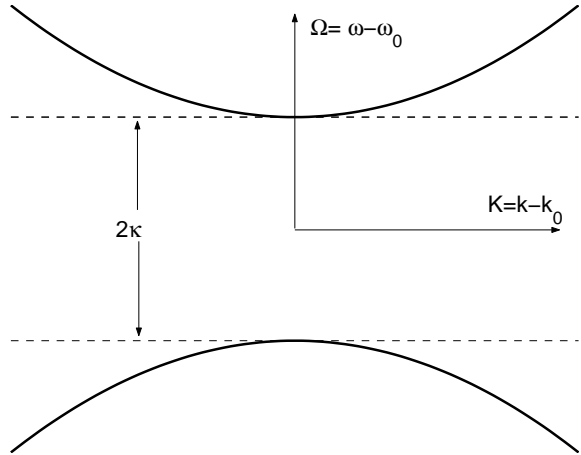}
\caption{The dispersion relation for the linearized coupled mode equations,
showing the spectral gap.}
\label{fig:gap}
\end{center}\end{figure}

 The opening of the first ``photonic band gap'' can be deduced from 
\eqref{eq:linearcme}. Seeking solutions to~\eqref{eq:linearcme} of the form
\begin{equation}
\label{eq:cmepw}
\binom{\Eplus}{\Eminus} = 
   e^{i(QZ-\Omega(Q) T)} \binom{\cE_+}{\cE_-}\enspace ,
\end{equation}
we find
\begin{equation}
\label{eq:OmegaQ}
\Omega^2(Q) = n^{-2} Q^2 + \k^2\enspace.
\end{equation}
The photonic band gap is pictured in Figure~\ref{fig:gap} which clearly shows
a region of excluded frequencies centered around $\Omega =0$.  For $\Omega$ in
the gap, $Q$ is imaginary, indicating that frequencies in Bragg resonance with
the grating cannot propagate.

Finally, combining~\eqref{eq:cmepw} and~\eqref{eq:OmegaQ}
with~\eqref{eq:cm-ansatz} gives the following approximation to a {\it band
edge} Floquet-Bloch generalized plane wave:
\begin{multline}
E(z,t;K)\left|_{K=k_B+\e Q}\right.\\
= \cE_+ e^{i[ (k_B - \e Q)z - (\w_B + \e\Omega(Q))t ]}
   + \cE_- e^{-i[ (k_B + \e Q)z + (\w_B + \e\Omega(Q))t ]}
+ \order(\e)\enspace.
\end{multline}

\subsection{Nonlinearity, instantaneous response and no periodic structure}
\label{sec:aperiodic_nl} 
Here, we take $\w_0 \to \infty$ and $\e=0$ in~\eqref{eq:AMLE}. The equations
then reduce to:
\begin{subequations}
\label{eq:quasilinear}
\begin{equation}
\pt^2 (E+ P)= \pz^2 E \enspace;
\end{equation}
and, for small $E$, 
\begin{equation}
P = P(E)= (n^2-1) E + \chi^{(3)}E^3 + \dots\ \enspace. \label{eq:quasilinear2}
\end{equation}
\end{subequations}
These may be combined to give
\begin{subequations} \label{eq:shocker}
\begin{align}
\pt^2 D(E) &= \pz^2 E\\
 D(E) &= n^2 E +  \chi^{(3)}E^3+ \dots \enspace.
\end{align}
\end{subequations}

To study this system we first rewrite it in a more standard form.  Let
$v\equiv {\cal K}^{-1}(E) \equiv D(E).$ Then,~\eqref{eq:shocker} becomes
$\pt^2 v = \pz^2{\cal K}(v)$.  Introducing $\pz u = v$, we obtain after one
integration
\begin{equation} \label{eq:nlstring}
\pt^2 u = \pz {\cal K}(\pz u)\enspace.
\end{equation}
Equation~\eqref{eq:nlstring} has the form of the equation governing the
vibrations of a nonlinear string, where the  electric displacement, $D(E)$,
plays the role of the strain, $\pz u$.\footnote{The classical relation
between tension, $\tau$ and strain $\pz u$, is derived via 
$${\cal K}(\pz u)= \pz u(1+(\pz u)^2)^{-\half}
\tau(\pz u).$$} 

A classical result of Lax~\cite{L:73} states that systems which are {\em
genuinely nonlinear} in the sense that ${\cal K}''(0)\ne0,$ or equivalently
$D''(0)\ne0$, will develop singularities in finite time.  Since $D''(0)=0$,
the quasilinear~\eqref{eq:shocker} does not satisfy the genuine nonlinearity
condition, although $D'''(0)\ne0$.  Klainerman and Majda~\cite{KM:80}
generalized Lax's result; if $\partial_E^{(p+1)}D(0)\ne0$ and the
initial data is of size $\e$ then singularity formation takes place within a
time interval of length $\order(\e^{-p})$.

In particular, it follows from this result that for initial data of size
$\order(\sqrt{\e})$ (see~\eqref{eq:lead}), $u(z,t)$ develops a singularity in
its second derivatives within $\order(\e^{-1})$ time.  Thus, $E$ remains
bounded but $\pz E$ tends to infinity at the singularity time. This is a {\it
shock} type singularity.  Specifically the results of~\cite{KM:80} imply the
following:
\begin{thm}
Consider the quasilinear wave equation~\eqref{eq:shocker} with smooth initial
data $E(z,t=0)$, $\pt E(z,t=0)$ which are of order $\sqrt{\e}$. Then, there
exists a finite and positive time, $T(\e)\le C\e^{-1}$ such that
\begin{equation}
\sup_{0\le t\le T(\e)} ||E(\cdot,t)||_\infty <\infty \enspace ,
\end{equation}
while
\begin{equation}
\lim_{ t\nearrow T(\e)} ||\pz E(\cdot,t)||_\infty
+||\pt E(\cdot,t)||_\infty=\infty\enspace.
\end{equation}
\label{thm:shock}
\end{thm}
Such carrier shock formation in the context of nonlinear optics has been
discussed by heuristic arguments in~\cite{FPM:96,GMR:99}.

\begin{figure}\begin{center}
\includegraphics[width=4in]{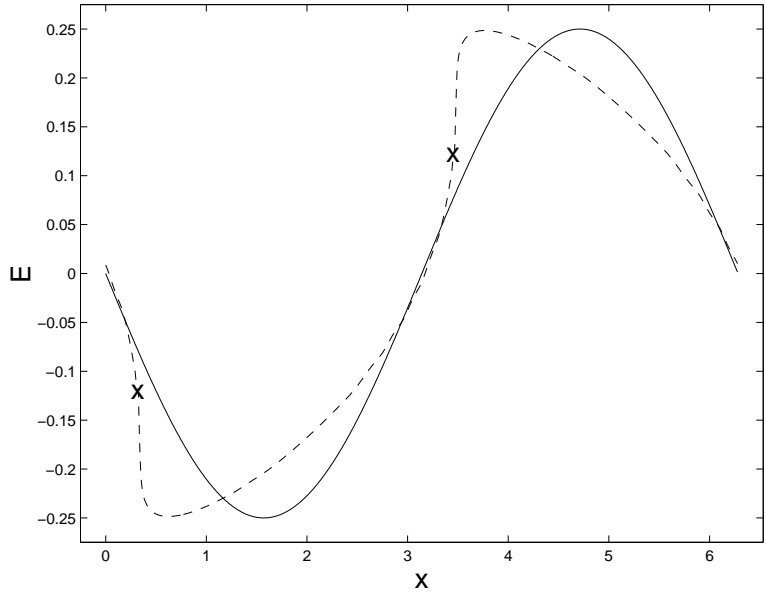}
\caption{Evolution from sinusoidal initial conditions (solid line) to near
shock formation (dashed line, with shock location at the {\bf x}) in
Maxwell's equation with instantaneous nonlinear polarization.  } 
\label{fig:shock}
\end{center}\end{figure}

Figure~\ref{fig:shock} shows a simulation of the shocking process on a single
carrier wave.\footnote{This computation is actually performed on the AMLE
system with a very fast material response $\w_0 \gg1$; see
Appendix~\ref{sec:comp_details}.  The simulation is run to the time the shock
``would have formed'' in the absence of material dispersion.} Computational
details are given in Sect.~\ref{sec:numerics}, and computational parameters
for this and all subsequent numerical results are given in
Appendix~\ref{sec:comp_details}.

\subsection{Nonlinearity, finite time response and no periodic structure:} 
\label{sec:nl_finite_time}
The mathematical model in this case is AMLE,~\eqref{eq:AMLE} with $\e=0$.
Joly, Metivier, and Rauch~\cite{JMR:96} proved that the initial value problem
for the full three dimensional AMLE, for some class of nonlinearities, does
not develop singularities in finite time.  In Sect.~\ref{sec:ivp} we outline a
proof of this result for our simpler one-dimensional model. Therefore,
material dispersion, resulting from the finite response time
($\w_0<\infty$), inhibits shock formation by providing a mechanism for
expelling high frequency modes away from the steepening regions. 

Numerical experiments suggest an interesting small dispersion limit as $\w_0$
tends to infinity in~\eqref{eq:AMLE2}. Note that for $0<\w_0 < \infty$ the
system is semilinear, but the limiting system is quasi-linear.

\begin{figure}\begin{center}
\includegraphics[width=4in]{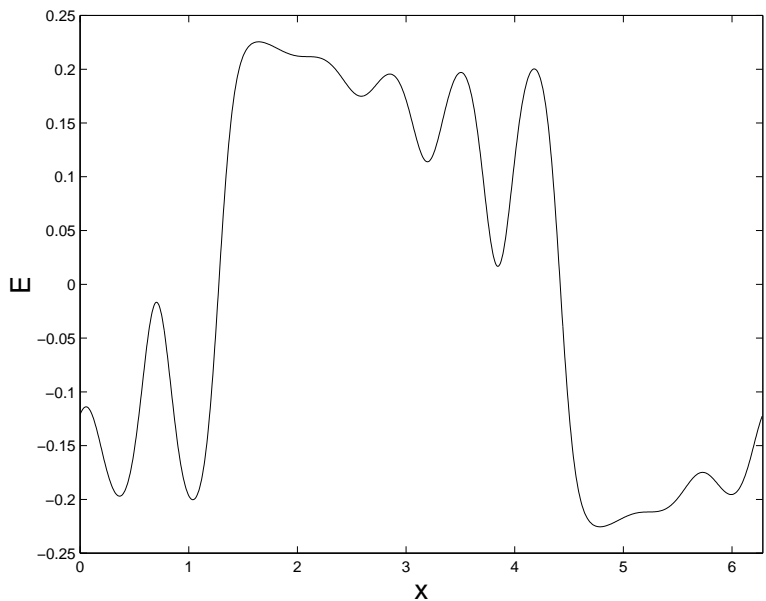}
\caption{Solution of AMLE after the ``shock time'' with small $\w_0$. } 
\label{fig:small_omega}
\end{center}\end{figure}

\begin{figure}\begin{center}
\includegraphics[width=4in]{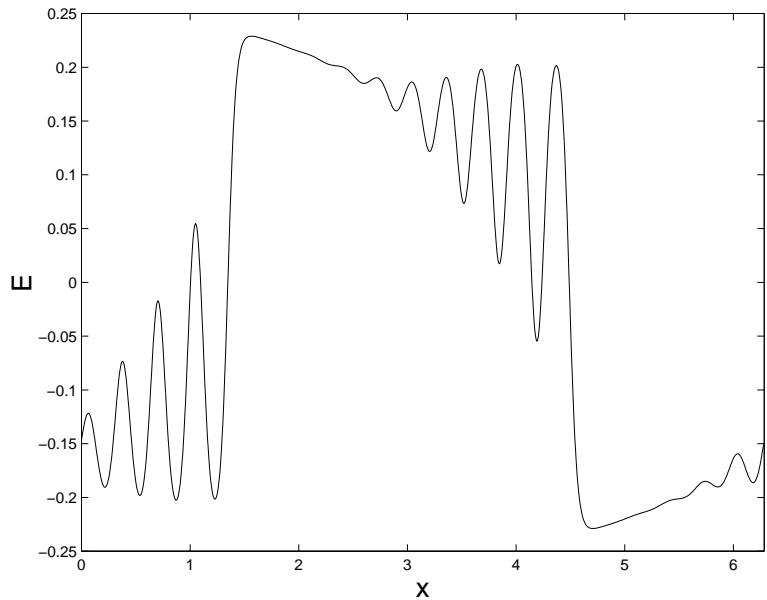}
\caption{Solution of AMLE after the ``shock time'' with $\w_0$ twice that of
Figure~\ref{fig:small_omega}.}  
\label{fig:big_omega}
\end{center}\end{figure}

Two computations with increasing values of $\w_0$ are shown in
Figures~\ref{fig:small_omega} and~\ref{fig:big_omega} at a short time after
the shock formation time in the dispersionless ($\w_0=\infty$) limit.  As
$\w_0 \to \infty $ the number of oscillations increases and in some weak
sense, the solution more closely approximates a weak solution to the Maxwell
system with instantaneous nonlinear polarization. 

The small material dispersion ($\w_0\gg1$)  limit of AMLE is analogous
to the  small dispersion limit of the Korteweg-de Vries equation (KdV), 
\begin{equation}
\pt u + u\pz u + \varepsilon\pz^3u=0\enspace ,
\label{eq:kdv}
\end{equation}
the equation of the free surface of an air/water interface in the regime of
long waves of small amplitude. The dispersionless $\varepsilon=0$ equation is
the {\it inviscid Burgers equation} and is easily seen to develop shocks
(singularities in derivatives of $u$) in finite time~\cite{W:74}. For
$\varepsilon\ne0$ solutions of KdV do not develop
singularities~\cite{K:81}. For initial data which give rise to shock formation
for $\varepsilon=0$, one observes, for $\varepsilon$ small, a scenario
analogous to what we observe for AMLE in the limit of $\w_0$ large.  KdV is an
integrable Hamiltonian system which is exactly solvable using the inverse
scattering transform (IST)~\cite{GGKM:67}. IST was used by Lax and
Levermore~\cite{LL:83} and by Venakides~\cite{V:85} to study this small
dispersion limit. In particular, the generation of oscillations is related to
the dynamics of solitons.  As in the case of KdV, for AMLE one observes the
generation of solitary wave like oscillations as a result of carrier wave
steepening. Computer simulations indicate that these solitary waves interact
more strongly and generate radiation, a manifestation of AMLE's apparent
nonintegrability.

\subsection{Periodic structure with nonlinearity, instantaneous response} 
\label{sec:periodic_nl}
In this case, the electric field is governed by
\begin{subequations}
\label{eq:pni}
\begin{align}
\pt^2 D(E,z) &= \pz^2 E\enspace ;\\
D(E,z) &= \left( n^2 + 2\e\nu\cos(2k_Bz)\right)E +\phi E^3
\enspace. \label{eq:instant}
\end{align}
\end{subequations}

The multiple scales approach implemented in Sect.~\ref{sec:derivation}
formally yields an expansion of the form (compare with~\eqref{eq:lead})
\begin{equation}
 E = \sqrt{\e}\sum_{m\ge1}\left( E_+^{(m)}(Z,T)e^{imk_B(z-t/n)}
 \ +\  E_-^{(m)}(Z,T)e^{-imk_B(z+t/n)} \right) + {\rm c.c.}\ +\ 
\e^{\frac{3}{2}} E_1
\enspace ,
\label{eq:infinitesum}
\end{equation}
where $E_\pm^{(m)} = E_\pm^{(m)}(Z,T) \equiv E_\pm^{(m)}(\e z,\e t),\ m\ge1$
satisfies a coupled system of infinitely many partial differential
equations. This is in contrast to the case $\w_0<\infty$, where the expansion
is replaced by~\eqref{eq:lead} involving only the {\it two} amplitudes
$E_\pm^{(1)}$ at leading ($\order(\sqrt{\e})$) order.

The reason for this difference can be seen by examining the equation for the 
correction, $E_1$, which takes the form:
\begin{equation}\begin{split}
\left(n^2 \pt^2 - \pz^2 \right) E_1 &= 
\sum_{q\ge1} [\ A_q^+(T,Z) e^{iq(kz-\w(k) t)}\ +\ 
 A_q^-(T,Z) e^{iq(kz+\w(k)t)}\ ]\\
&= \sum_{q\ge1} [\ A_q^+(T,Z) e^{iqk(z-t/n)}\ +\ 
 A_q^-(T,Z) e^{iqk(z+t/n)}\ ] \enspace.
\label{eq:infinite_system}
\end{split}\end{equation}
The coefficients, $A_q^\pm$ involve the unknown amplitudes $E_\pm^{(m)}$ and
their derivatives. In order for $\e E_1$ to be smaller than the first term in
the expansion of $E$, it is necessary to remove all resonances from the right
hand side. Resonances are excited by components of the right hand side which
are plane waves of the homogeneous problem. If $0<\w_0<\infty$, the
unperturbed dynamics are dispersive ($\w(qk)\ne q\w(k)$).  Therefore, the
contributions to the above sum for $q\ge2$ are nonresonant and the
nonresonance condition implies coupled equations for $E^{(1)}_+$ and
$E^{(1)}_-$ alone. In the case of instantaneous response ($\w_0=\infty$:
absence of material dispersion) all terms in the sum are resonant.  Therefore,
in order to preclude secular growth, we require $A_m^\pm\equiv 0,\ m\ge1$.
This yields a coupled system of {\it infinitely} many equations governing the
evolution of the backward and forward amplitudes: $E_\pm^{(m)}(Z,T),\
m\ge1$. We do not address the question of whether the approximate solution
generated, via~\eqref{eq:infinitesum}, is a convergent series which represents
an approximate solution of Maxwell's equation.
 
Indeed, the contrast we find between the dispersive ($\w_0$ positive and
finite) and nondispersive ($\w_0=\infty$) cases is consistent with the
observations of the previous section concerning shock formation and therefore
the generation of high frequency harmonics.\footnote{ Note, however, that the
simulations described in the previous section are not for wave-packet initial
conditions. It is reasonable to ask whether the dispersion which comes from
the photonic band structure is sufficient to regularize shocks. Preliminary
direct simulations for the nondispersive ($\w_0=\infty$) limit employing
numerical schemes designed to capture shock-like structures indicate that
shocks very likely form in the carrier, though the {\it envelope} appears to
evolve smoothly (E. Kirr, in progress).}

Consequently, the NLCME system does not describe the evolution of the wave
packet envelope for the system without material dispersion. Although the ratio
of the effects of photonic band dispersion to material dispersion in the
experiments of Eggleton, Slusher et. al.~\cite{E:97,E:99,E:96} is of order
$10^6$, we argue that nonlinearity rapidly (on a time scale of order
$\tilde\w_0^{-1}$) generates frequency content for which material dispersion
is significant; see the appendix. As noted in~\ref{sec:nl_finite_time},
material dispersion regularizes the wave steepening by propagating nonlinearly
generated frequencies, which are nearly resonant, away from a steepening
front.

\subsection{ Periodic structure, nonlinearity and finite time response}
\label{sec:periodic_nl_finite}
In this case, we have the full AMLE equations~\eqref{eq:AMLE}. We show in
Theorem~\ref{thm:main} that for small amplitude waves in the SVEA regime,
solutions to AMLE are well approximated by solutions to the Nonlinear Coupled
Mode Equations (NLCME).

The NLCME have a well-known class of solutions known as ``gap solitons'' which
are able to propagate through the fibers at any velocity between zero and the
speed of light.  We present them in the general form as derived by Aceves and
Wabnitz~\cite{AW:89}.  The solutions depend on two parameters, $\abs{v}<1$ and
$\d$:
\begin{subequations}
\label{eq:gapsoliton}
\begin{align}
\Eplus &= s \alpha e^{i\eta} \sqrt{\Bigl \lvert\frac{\k}{2\G}\Bigr \rvert}
\frac{1}{\Delta}
\left(\sin{\d}\right) e^{i s \sigma} \sech{(\theta - i s\d/2)}\enspace;\\
\Eminus &= - \alpha e^{i\eta} \sqrt{\Bigl \lvert\frac{\k}{2\G}\Bigr\rvert}
\Delta   
\left(\sin{\d}\right) e^{i s \sigma} \sech{(\theta + i s\d/2)}\enspace;
\end{align}
\end{subequations}
where:
\begin{gather*}
\gamma = \frac{1}{\sqrt{1-v^2}} \enspace ; \qquad
\Delta = \left(\frac{1-v}{1+v}\right)^{\frac{1}{4}}\enspace ; \\
\theta = \g \k (\sin{\d})(v_g^{-1}Z-v T)\enspace ; \qquad
\sigma = \g \k (\cos{\d})(v_g^{-1} v Z - T)\enspace ; \\
s = \sign{(\k \G)}\enspace ; \qquad
\alpha = \sqrt{\frac{2(1-v^2)}{3-v^2}}\enspace ;  \\
e^{i\eta}= \left( - \frac
{e^{2\theta} + e^{- i s \d}}
{e^{2\theta} + e^{i s\d}} \right)^{\frac{2v}{3-v^2}}.
\end{gather*}

Combining this family of exact solutions to NLCME with Theorem~\ref{thm:main}, we have the following corollary:
\begin{cor}
The gap solitons approximate to $\order(\e)$ a family of long-lived   solutions
to the AMLE system for times of $\order(\e^{-1})$.
\end{cor}

The gap solitons solutions bear a striking resemblance to solitons of the
Nonlinear Schr\"odinger equation (NLS).  In fact, in the limit $\delta \ll 1$,
$v \ll 1$, we may show that the gap soliton may be written as a normal mode of
linear coupled mode equations, slowly modulated by an NLS soliton.  To see
this, we study the NLCME themselves under the SVEA limit.  We assume $\delta$
small, and look for approximate solutions to~\eqref{eq:cme} of the form:
\begin{equation}
\binom{\Eplus}{\Eminus} \approx  \delta A(\d Z, \d T, \d^2 T) \Vec V 
e^{i(QZ-\Omega T)}
\label{eq:nlcmetonls}
\end{equation}
where $\Vec V e^{i(QZ-\Omega T)}$ solves the linearized NLCME.
Then $\Omega_\pm = \pm \sqrt{\k^2 + v_g^2 Q^2}$ and $A$ solves
\begin{equation}
 i \partial_\tau A + \frac{\Omega''(Q)}{2} \partial_\z^2 A + N
 \abs{A}^2 A
\end{equation}
with 
\begin{gather}
\z = \d (Z - \Omega'(Q) T)\enspace, \qquad 
\tau = \d^2 T \enspace, \nonumber \\
\text{and }  N = \frac{\G}{2}\left(3-\frac{v_g^2 Q^2}{\k^2 + v_g^2
Q^2}\right) \enspace. \nonumber
\end{gather}

NLS has spatially localized standing wave solutions of the form
\begin{equation}
A(\z,\tau)=
\pm \sqrt{\frac{2 \lambda}{N}} e^{i\lambda \tau}
\sech{\left(\sqrt{\frac{2\lambda}{\Omega_+''(Q)}} \z\right)}
\end{equation}
and if we let $Q=0$ and $\lambda = \frac{\kappa}{2}$ in this formula, then we
recover exactly the leading term in the expansion of the gap soliton for $v=0$
and $\delta \ll 1$.  De Sterke and Sipe show additionally that the first two
terms in the expansion of the gap soliton for small $\delta$ and $v$
correspond to the $\order(\d)$ and $\order(\d^2)$ terms in the multiple scales
construction of solutions to AMLE~\cite{DS:94} with small wavenumber $Q$. 

Therefore, we expect the following relationship among AMLE, NLCME and NLS. For
$\delta<\delta_0$ sufficiently small NLCME has a solution of the
type~\eqref{eq:nlcmetonls}, where $A(\z,\tau)$ satisfies NLS.  The validity
of NLS as an approximation to NLCME could be shown using the methods presented
in Sect.~\ref{sec:validity} and~\cite{KSM:92}.  This solution of NLCME,
generated by NLS, gives rise to a solution of AMLE of the
type~\eqref{eq:lead}, provided $\e\le\e(\delta_0)$ is sufficiently small.

\section{Derivation of the Nonlinear Coupled Mode Equations}
\label{sec:derivation}
In this section we use the method of multiple scales~\cite{W:74} to derive the
nonlinear coupled mode equations.  We begin with the equation:
\begin{subequations}
\begin{equation}
 \pt^2\left( E +P\right)= \pz^2E \enspace. \label{eq:M1} 
\end{equation}
We also specify a more general form for the nonlinear response in
modeling the polarization in~\eqref{eq:AMLE2}:
\begin{equation}
\w_0^{-2}\pt^2 P + (1-2\e \nu \cos(2k_Bz))P + g(P,z)  = (n^2-1)E \enspace,
\label{eq:M3}
\end{equation}
\end{subequations}
where, for small values of $P$, 
\begin{equation}
g(P,z) = -\phi P^3 + {\rm higher\ order\ terms}.
\end{equation}
We expand the dependent variables in powers of $\e$:
\begin{align}
E &= \e^{\half}E_0 + \e^{\frac{3}{2}}E_1 + \e^{\frac{5}{2}}E_2 + \dots
\label{eq:E_series}\\ 
P &=\e^{\half}P_0 + \e^{\frac{3}{2}}P_1 + \e^{\frac{5}{2}}P_2 + \dots
\end{align}
and expand the derivatives in terms of slow scales $Z=\e z$ and $T = \e t$:
\begin{equation}
 \pt \to \pt + \e \pT  \mbox{ and } \pz \to \pz + \e \pZ\enspace.
\end{equation}
To derive the NLCME, it will be sufficient for us to consider the equations
for the first two terms in the expansion, which may be written
\begin{align}
\order(\e^{1/2}):\;\cL_0 \binom{E_0}{P_0} &= 0 \enspace, \label{eq:eps1_2}\\
\order(\e^{3/2}):\;\cL_0 \binom{E_1}{P_1} &= -\cL_1 \binom{E_0}{P_0} +\binom{0}
{2\nu\cos(2k_Bz)P_0+\phi P_0^3}\label{eq:eps3_2} \enspace,
\end{align}
where
\begin{equation}
 \cL_0 = \cL_0( i^{-1}\pt,i^{-1}\pz)= 
 \begin{pmatrix}   
  \pt^2 -\pz^2 & \pt^2 \\ 
  1-n^2 & 1 +\w_0^{-2}\pt^2 
 \end{pmatrix}
 \enspace ,
 \label{eq:L0}
\end{equation}
and
\begin{equation}
\cL_1 = 2\begin{pmatrix}   \pt \pT -\pz \pZ & \pt\pT \\ 0 &
 \w_0^{-2}\pt\pT
\end{pmatrix}\enspace.
\end{equation}
We now seek solutions order by order.
\subsubsection{$\order(\e^{1/2})$}

At this order, the solution to the linear problem is given by~\eqref{eq:E0P0},
where $k$ and $\w$ satisfy dispersion relation~\eqref{eq:disp_rel}.  As $k$
will be determined by the length scale of the Bragg grating structure, we
prefer to solve~\eqref{eq:disp_rel} for $\w$ as a function of $k$:
\begin{equation}
\w^2 = \half\left( n^2 \w_0^2+ k^2  \right) \pm \half \sqrt{ (n^2 \w_0^2
+k^2)^2 -4 \w_0^2 k^2} \label{eq:omega}\enspace.
\end{equation}
Equation~\eqref{eq:omega} has two roots corresponding to each choice of sign.
In the limit as $\w_0 \to \infty$, the root corresponding to the plus sign
diverges to $\infty$, while the root corresponding to the minus sign
approaches the finite value $n^2\w^2 = k^2$, as noted in
Sect.~\ref{sec:linear_homogeneous}.  We examine a pair of backward and forward
propagating modes in Bragg resonance with the fiber and having slowly varying
amplitudes:
\begin{equation}
\binom{E_0}{P_0} = \left( \Eplus(Z,T) e^{i(k_Bz-\w_B t)} + 
\Eminus(Z,T) e^{-i(k_Bz+\w_B t)} + c.c.\right)\ \binom{1}{\g_B} \enspace,
\label{eq:def_E0P0}
\end{equation}
where 
\begin{align}
k_B&=\frac{\pi}{d}\enspace ,\\
\gamma_B &= \frac{n^2-1}{1-(\frac{\w_B}{\w_0})^2}
\enspace ,
\label{eq:gb}
\end{align}
and $\w_B$ is a root of~\eqref{eq:omega} for the minus sign choice. 

\subsubsection{$\order(\e^{3/2})$}
The equation at this order is
\begin{equation}
\cL_0 \binom{E_1}{P_1} = - \cL_1\binom{E_0}{P_0} 
+\binom{0}{2 \nu \cos(2k_Bz) P_0 + \phi P_0^3} \enspace.
\end{equation}
Substituting in the solution to the $\order(\e^{1/2})$ equation, we find
\begin{equation}\begin{split}
\cL_0 \binom{E_1}{P_1} = &  
\binom{ 2i\left(k_B\pZ \Eplus  + \w_B(\g_B +1)\pT\Eplus\right)} 
{\frac{2i\g_B \w_B}{\w_0^2}\pT \Eplus + \g_B \nu\Eminus +  
  3\g_B^3\phi(\abs{\Eplus}^2 + 2\abs{\Eminus}^2 )\Eplus}  
e^{i(k_Bz-\w_B t)} \\
&+\binom{2i\left(-k_B\pZ \Eminus + \w_B(\g_B +1)\pT \Eminus\right)} 
 {\frac{2i\g_B
\w_B}{\w_0^2}\pT \Eminus + \g_B\nu \Eplus +  3 \g_B^3\phi(\abs{\Eminus}^2 +
2\abs{\Eplus}^2 )\Eminus  }  e^{-i(k_Bz+\w_B t)} \\ 
&+ \binom{0}{\g_B^3\phi\Eplus^3 }e^{3i(k_Bz-\w_B t)}\\ 
&+ \binom{0}{3\g_B^3\phi\Eplus^2\Eminus}e^{i(k_Bz-3\w_B t)}\\ 
&+ \binom{0}{3\g_B^3\phi\Eplus\Eminus^2} e^{-i(k_Bz+3\w_B t)}\\
&+ \binom{0}{\g_B^3\phi\Eminus^3} e^{-3i(k_Bz+\w_B t)}\\ 
&+ \binom{0}{3\g_B^3\phi
\Eplus^2 \Eminus^* +\g_B \nu \Eplus } e^{i(3k_Bz-\w_B t)}\\ 
&+\binom{0}{3\g_B^3 \phi
\Eminus^2 \Eplus^* +\g_B \nu\Eminus} e^{-i(3k_Bz+\w_B t)} +c.c.\enspace.
\label{eq:bigugly}\end{split}\end{equation}

Of the terms on the right hand side of~\eqref{eq:bigugly}, only the first two
are potentially resonant, and may therefore give rise to secular growth in
time, $t$. The nonresonance condition required to remove such resonances can
be expressed as the requirement that the vector coefficients of
$e^{i(k_B-\w_Bt)}$ and $e^{-i(k_Bz+\w_B t)}$ both lie in the column space of
$\cL_0(\w_B,\pm k_B)$; see~\eqref{eq:L0}.  Equivalently, we require that the
inner product of each of these vectors with the vector
$(-{\cL_0}_{2,1},{\cL_0}_{1,1})$ be equal to zero. This yields the Nonlinear
Coupled Mode Equations (NLCME):
\begin{subequations}
\label{eq:nlcme}
\begin{align}
i \left( \pT \Eplus + v_g \pZ \Eplus \right) + \k \Eminus +
\G(\abs{\Eplus}^2 + 2\abs{\Eminus}^2)\Eplus &= 0 
\enspace;\label{eq:nlcme1}\\ 
i\left(  \pT \Eminus - v_g \pZ \Eminus \right) + \k \Eplus +
\G(\abs{\Eminus}^2 + 2\abs{\Eplus}^2)\Eminus &= 0
\enspace. \label{eq:nlcme2}
\end{align}
\end{subequations}
Here
\begin{equation}
v_g  = \w'(k_B) = k'(\w_B)^{-1} =
\frac{k_B(\frac{\w_B^2}{\w_0^2}-1)}
{\w_B\left( \frac{\w_B^2}{\w_0^2}- \left(1+\g_B\right)\right)}
\label{eq:cg}
\end{equation}
is the group velocity, and the coupling and nonlinearity parameters are 
\begin{equation}
\k = \frac{\w_B(n^2-1)}{2\left(n^2-1 + \left(1- \frac{\w_B^2}{\w_0^2}\right)^2
\right)} \nu, \qquad 
\G = \frac{3\g_B^3\w_B^2}{k_B(1-\frac{\w_B^2}{\w_0^2})}\ \phi \enspace.
\label{eq:kgamma}
\end{equation}

Our proof of validity of NLCME on time scales of order $\e^{-1}$ requires that
we solve explicitly for $E$ and $P$ through order $\e$.  We
solve~\eqref{eq:bigugly} and obtain
\begin{equation}
\binom{E_1}{P_1}= \sum_{\substack{ a= \pm 1, \pm 3 \\ b = 1,3}}
\binom{E_1^{(a,b)}}{P_1^{(a,b)}}+c.c.
\end{equation}  
such that for $(a,b) \neq
(\pm 1,1)$,
\begin{equation}
\cL_0\binom{E_1^{(a,b)}}{P_1^{(a,b)}} = \binom{0}{S^{(a,b)}} e^{i(ak_Bz-b\w_B t)}\enspace ,
\end{equation}
where $S^{(a,b)}$ is determined in equation~\eqref{eq:bigugly}, and for $(a,b)
= (\pm 1,1)$, the right hand side is determined by using~\eqref{eq:nlcme} to
eliminate $\pT E_{\pm}$ from the first two terms of~\eqref{eq:bigugly}.  Once
this is done, these terms take the form of the non-null eigenvectors of
$\cL_0$ and solving this part of the equation becomes a trivial linear algebra
problem.  In this way we may represent the approximate solution using only
$E_\pm$ and their $Z$-derivatives, so that $L^2$ and $H^s$ estimates on
solutions to the NLCME suffice for proof of the main theorem.

\section{ The initial value problem for AMLE and NLCME}
\label{sec:ivp}
Our proof of the validity of NLCME as an approximation to AMLE requires some a
priori knowledge of the solutions of these equations. In this section we
outline the theory of the initial value problems for AMLE and NLCME and
collect the necessary facts for the proof of the main theorem.

Both AMLE and NLCME are semilinear hyperbolic systems whose initial value
problems can be expressed in the form:
\begin{equation}\begin{split}
\label{eq:semilin}
\pt \Phi(t) &= -i A \Phi(t) + J[\Phi(t)]\enspace,\\
\Phi(t=0) &= \Phi_0\enspace.
\end{split}\end{equation}
Here, $A$ is a self-adjoint operator on a Hilbert space ${\cal H}$ and 
$J$ is a nonlinear mapping from ${\cal H}$ to itself and $\Phi_0\in {\cal H}$.

We first indicate how AMLE and NLCME can be expressed in this form and then
show how the general theory and {\it energy estimates} can be used to conclude 
the existence of global in time solutions.
\medskip

\noindent{\bf AMLE:}

To write the AMLE system,~\eqref{eq:AMLE}, as a first order system we use the
variables:  $E,B,P$ and $Q\equiv\pt P$. The AMLE system then becomes:
\begin{equation}\label{eq:EBPQ}
\begin{split}
\pt E&= \pz B - Q \enspace ; \\
\pt B&= \pz E \enspace ; \\
\pt P&= Q \enspace ;\\
\pt Q&= -\w_0^2(1-2\e \nu\cos(2k_Bz))P-\w_0^2g(P,z)+\w_0^2(n^2-1)E
\enspace.
\end{split}
\end{equation}

We now write~\eqref{eq:EBPQ} in a more compact form. Let
\begin{equation}
\vu = \begin{pmatrix} E \\ B \\ P \\Q \end{pmatrix},
 \ {\rm and}\ {\cal M} =
\begin{pmatrix}
0 & \pz & 0 & -1 \\ \pz & 0 & 0 & 0 \\ 0 & 0 & 0 & 1 \\
 \w_0^2(n^2-1) & 0 &
-\w_0^2& 0
\end{pmatrix} \enspace.
\end{equation}
Then, the full system may be written
\begin{equation}
\pt \vu = {\cal M}\vu\ + \w_0^2\efour
 \left(2\e\nu \cos(2k_Bz) P - g (P,z) \right),
\label{eq:compact}
\end{equation}
where $\efour=(0,0,0,1)^T$.
Thus, 
\begin{align}
\Phi&=\vu,\ A = i{\cal M} \enspace, \\ 
J[\Phi]&=\w_0^2\efour\left(2\e\nu \cos(2k_Bz) P - g (P,z) \right)
\enspace.
\end{align}
\medskip

\noindent{\bf NLCME:}

NLCME can be written in the form~\eqref{eq:semilin} with the definitions:
\begin{align}
\label{eq:nlcme-notation}
\Phi &=\ \binom{E_+}{E_-}\enspace ; \\ 
\sigma^1 & =  \begin{pmatrix} 0 & 1\\ 1 & 0\end{pmatrix} \enspace ;\\
A &= -iv_g\pZ - \k\sigma^1 \enspace ;\\
J[\Phi] &=  i \G \binom
{\left(\abs{\Eplus}^2 + 2 \abs{\Eminus}^2\right) \Eplus}
{\left(\abs{\Eminus}^2 + 2 \abs{\Eplus}^2\right) \Eminus}
\enspace.
\end{align}

We now formulate the general initial value problem~\eqref{eq:semilin} as an
equivalent integral equation: 
\begin{equation}
\Phi(t) = e^{-iAt}\Phi_0 +\int_0^t e^{-iA(t-s)} J[\Phi(s)] ds\enspace.
\label{eq:inteqn}
\end{equation}

It is elementary to show~\cite{Reed:76} using the contraction mapping
principle that in both examples, for any initial condition, $\Phi_0$, in the
Hilbert space ${\cal H}=H^1$, there is a maximal time,
$T_{max}=T_{max}(||\Phi_0||_{H^1})>0$, and a solution $\Phi(t)$
of~\eqref{eq:inteqn}, which is defined for $t\in [0,T_{max})$, the maximal
time interval of existence.  The solution $\Phi(t)\in H^1$ for each $t\in
[0,T_{max})$ and the function $t\mapsto ||\Phi(t)||_{H^1}$ is continuous for
$t\in [0,T_{max})$.  Finally, either $T_{max}<\infty$ or $T_{max}=\infty$. If
$T_{max}<\infty$, then
\begin{equation}
\label{eq:blowup}
\lim_{t\nearrow T_{max}} ||\Phi(t)||_{H^1}=\infty\enspace.
\end{equation}
and we say that the solution $\Phi(t)$ blows up at time $T_{max}$ in $H^1$.
As we have seen in Theorem~\ref{thm:shock} of Sect.~3, in the absence of
material dispersion, solutions of the AMLE system do develop singularities in
their gradients in finite time. We claim that for both dispersive systems AMLE
and NLCME no singularities form:

\begin{thm}
\label{thm:global}
For initial data in $H^1$, $T_{max}=\infty$. That is, AMLE (under
Hypothesis~\ref{hyp:N1} below on the nonlinearity) and NLCME have global in
time $H^1$ solutions.
\end{thm}

To prove this theorem, it suffices to show that if $T_1$ is an arbitrary time,
then the $H^1$ norm of any of the components of $\Phi(t)$ satisfies an
estimate:  
\begin{equation}
\label{eq:h1bound}
||\Phi_j(t)||_{H^1} \le C(T_1)\enspace.
\end{equation}
The constant, $C(T_1)$ may depend on and even grow with $T_1$, but must be
finite for finite values of $T_1$. To prove~\eqref{eq:h1bound} we use a
combination of the conservation laws associated with AMLE and NLCME as well as
direct a priori estimates on the evolution equations. We consider the cases of
AMLE and NLCME individually.
\vspace{3mm}

\noindent{\bf Proof of Theorem~\ref{thm:global} for AMLE:}\\
We use the formulation for AMLE given in~\eqref{eq:compact} or
equivalently~\eqref{eq:EBPQ}.  Our proof makes use of the following technical
assumption on the nonlinear term which ensures the existence of global
solutions for arbitrary size $H^1$ data:
\medskip

\begin{HN}
There exists a constant $C$, such that for all $z$, 
\begin{equation}
|g(P,z)| + |\pz g(P,z)| \le C|P| \enspace,   
\quad \abs{\partial_P g(P,z)} \le C \enspace .
\end{equation}
\label{hyp:N1}
\end{HN}

The first step is to derive an {\it energy estimate} for AMLE. Taking the dot
product of~\eqref{eq:compact} with the vector 
$(E,B,(n^2-1)^{-1}P,\w_0^{-2}(n^2-1)^{-1}Q)$ yields:
\begin{equation}
\label{eq:energyestimate}
\begin{split}
& \diff{}{t} \half\int \left( E^2 + B^2 + 
\frac{1}{n^2-1}P^2 + \frac{1}{\w_0^2(n^2-1)}Q^2 \right) dz \\
&= \frac{2\e\nu}{n^2-1}\int\ \cos(2k_Bz)PQ\ dz\ 
 -\ \frac{1}{ n^2-1}\int g(P,z)Q \ dz\\
& \le C\int( P^2 + Q^2) dz
\enspace.
\end{split}
\end{equation}
The previous inequality follows from Hypothesis~\ref{hyp:N1}. It follows that
\begin{equation}
||\vu(t) ||^2_{L^2} \le ||\vu_0 ||^2_{L^2}  + 
C_1\int_0^t ||\vu(s) ||^2_{L^2}\ ds
\end{equation}
for some positive constant $C_1$ and therefore by Gronwall's inequality:
\begin{equation}
||\vu (t) ||_{L^2} \le  ||\vu_0 ||_{L^2}\  e^{C_1 t} \enspace.
\label{eq:l2estimate}
\end{equation}

Estimates for the $L^2$ norm of $\pz\vu$ are obtained in a similar manner.  We
first differentiate equation~\eqref{eq:compact} for $\vu$ with respect to $z$,
and then take the dot product with 
$(\pz E,\pz B,(n^2-1)^{-1}\pz P,\w_0^{-2}(n^2-1)^{-1}\pz Q)$ and obtain:
\begin{equation}
\begin{split}
&\diff{}{t}\half
\int\left( E_z^2+ B_z^2 + \frac{1}{n^2-1}P_z^2
+ \frac{1}{\w_0^2(n^2-1)}Q_z^2 \right)dz \\
&= \frac{\e\nu}{n^2-1} \int
     [\ 2\cos(2k_Bz)P_zQ_z-4k_B\sin(2k_Bz)PQ_z \ ]\ dz \\
&\ \ -\frac{1}{n^2-1}\int  \partial_Pg(P,z) P_z Q_z\ dz
 - \frac{1}{n^2-1}\int  \pz g(P,z) Q_z\ dz \\
&\le C \int        \left( P^2 + Q^2 + P_z^2 + Q_z^2 
                   \right)dz \enspace.
\end{split}
\end{equation}
This, together with the above $L^2$ energy estimate can be used to conclude,
by Gronwall's inequality, 
\begin{equation}
 ||\vu ||_{H^1} \le  ||\vu_0 ||_{H^1}\  e^{C_2 t}.
\end{equation}
Since the $H^1$ norm of $\vu$ grows at worst exponentially, we conclude that
$T_{max}=\infty$.  This completes the proof of $H^1$ existence for solutions
to AMLE.
\bigskip

\noindent{\bf Proof of Theorem~\ref{thm:global} for NLCME:}

\begin{prop}
Let $\vE=(\Eplus,\Eminus)$ satisfy system~\eqref{eq:nlcme} with initial
conditions  $\vE (0) \in H^s$ for $s\ge 1$.\footnote{In our proof of
Theorem~\ref{thm:main}, we use this result for $s\le3$.}  
Then there exists
$C_s = C_s\left(\norm{\vE(0)}{H^s},T\right)$ such that  
$\norm{\vE(T)}{H^s} \le C_s\left(\norm{\vE(0)}{H^s},T\right)$. 
Moreover, $C(x_1,x_2) \to 0$ as $x_1\to 0$\enspace.
\label{prop:E}
\end{prop}

\noindent{\bf Proof}\\
It is easy to see that system~\eqref{eq:nlcme} preserves the $L^2$ norm. To
obtain this and higher $L^p$ bounds on $\Epm$, we multiply both sides
of~\eqref{eq:nlcme1} by $\abs{\Eplus}^{2\sigma}\Eplus^*$ and~\eqref{eq:nlcme2}
by $\abs{\Eminus}^{2\sigma}\Eminus^*$, add them, and take the imaginary part,
yielding:
\begin{multline}
\frac{1}{\sigma+1} \bigl(
\pT(\abs{\Eminus}^{2\sigma+2} +\abs{\Eplus}^{2\sigma+2})\bigr)
+ v\pZ(\abs{\Eplus}^{2\sigma+2} -\abs{\Eminus}^{2\sigma+2}) \\
+i \G(\Eplus\Eminus^*+\Eminus\Eplus^*)
(\abs{\Eplus}^{2\sigma} -\abs{\Eminus}^{2\sigma})=0 \enspace.
\end{multline}
If $\sigma=0$, then the last term is identically zero, showing that
$\norm{\vE}{2}^2$  is conserved.%
\footnote{Recall that $\norm{\vE}{p}^p=\int
\left (\abs{\Eplus}^p + \abs{\Eminus}^p\right )dZ$.}
If $\sigma>0$, then we may bound $\norm{\vE}{}$ using Gronwall's inequality
\begin{equation}\begin{split}
\diff{}{T} \norm{\vE}{2\sigma+2}^{2\sigma+2} &\le  c
(\sigma +1)\norm{\vE}{2\sigma+2}^{2\sigma+2} \enspace ; \\
\norm{\vE}{2\sigma+2}^{2\sigma+2} &\le
\norm{\vE_0}{2\sigma+2}e^{c(\sigma+1)T}  \enspace ;\\
\norm{\vE}{2\sigma+2} &\le
\norm{\vE_0}{2\sigma+2}e^{cT/2} \enspace.
\end{split}\end{equation}
Letting $p=2\sigma +2$, this is just
\begin{equation}
\norm{\vE}{p} \le \norm{\vE_0}{p}e^{cT/2} \enspace.
\end{equation} 
As $c$ is independent of $p$, this estimate holds for $L^\infty$. 

The $L^\infty$ bound can then be used to bound growth rates of the $L^p$ norms
of $\pZ \Epm$ in terms of $T$.  Taking $Z$-derivatives of the NLCME, and
performing a similar calculation with $\sigma =0$ yields:
\begin{equation}\begin{split}
\diff{}{T} \norm{\pZ \vE}{2}^2
&\le c \norm{\vE \cdot \pZ \vE}{2}^2 \\
&\le c \norm{\vE}{L^\infty}^2 \norm{\pZ\vE}{L^2}^2 \\
&\le c \norm{\vE_0}{L^\infty}^2 e ^{2cT}\norm{\pZ\vE}{L^2}^2 \enspace,
\end{split}\end{equation}
so that
\begin{equation}
\norm{\vE}{H^1} \le \norm{\vE_0}{H^1} e^{c( e^{cT}-1)}\enspace.
\end{equation}
This shows that we can indeed bound the solutions of NLCME in $H^1$, and
control them for times $T=\order(1)$, i.e. $t=\order(\frac{1}{\e})$.
Proceeding similarly, we can derive bounds in higher Sobolev spaces,
specifically $H^2$ norms like $e^{ce^{ce^{cT}}}$, and $H^3$ bounds like
$e^{ce^{ce^{ce^{cT}}}}$, thus completing the proof of Prop.~\ref{prop:E}, and
hence, by the comments preceding Theorem~\ref{thm:global}, of that theorem.

\section{Validity of NLCME for times, $t$, of $\order(\e^{-1})$; proof of 
Theorem~1} 
\label{sec:validity}
We shall work with the formulation of AMLE given in~\eqref{eq:EBPQ}. 

We proceed under the following hypothesis concerning the nonlinearity
$g(P,z)$ and its derivative $\partial_P g(P,z)$ for small $P$:
\medskip

\begin{HN}
Assume $g$ has partial derivative of order $\le4$ with respect to $P$ and has 
one partial derivative with respect to $z$ which is continuous.  Assume
further that $g(0,z) = (\partial_Pg)(0,z)= (\partial_P^2g)(0,z)=0,$ and 
$\phi\equiv-\frac{1}{3!} (\partial_P^3g)(0,z)\ne0$ and is independent of $z$,
and make the analogous assumptions for $\pz g$ Therefore, there exists a
positive constant, $C$, such that for all $z$ and all $P_1, P_2$ with 
$|P_1|+ |P_2|$ sufficiently small:
\begin{equation}
\begin{split}
|g(P_1+P_2,z) - g(P_1,z)| &\le C \left( |P_1|^2+|P_2|^2 \right)
|P_2| \enspace ;\\ 
|\partial_Pg(P_1+P_2,z)-\partial_Pg(P_1,z)| 
&\le C \left( |P_1|+|P_2|\right)|P_2|\enspace ; \\
|\pz g(P_1+P_2,z) - \pz g(P_1,z)| 
&\le C \left( |P_1|^2+|P_2|^2 \right)|P_2| \enspace.\\ 
\end{split}
\end{equation}
\label{hyp:N2}
\end{HN}

To obtain an approximate solution of~\eqref{eq:EBPQ}, we require, in addition
to our approximations of $E$ and $P$, approximations to $B$ and $Q$ through
first order in~$\e$.  We use the relation 
$\pt B(t,z) = \pz E(t,z)$ to obtain the relations
\begin{align}
\pt B_0 &= \pz E_0 \enspace ; \label{eq:B0} \\
\pt B_1 &= -\pT B_0 + \pz E_1 + \pZ E_0 \enspace.
\end{align}
Also, using $Q=\pt P$, we find
\begin{align}
Q_0 &= \pt P_0 \enspace ; \label{eq:Q0}\\
Q_1 &= \pt P_1+\pT P_0\enspace.
\end{align}
We may then define 
\begin{equation}
X^\e_{app} = \e^{\half}\left(X_0 + \e X_1 \right) 
\mbox{ for }X=E, B, P, \mbox{ or }Q
\end{equation}
and write our approximate solution to~\eqref{eq:EBPQ} as:
\begin{equation}
\vu^\e_{app} = \begin{pmatrix} E^\e_{app} \\ B^\e_{app}\\ P^\e_{app} \\
Q^\e_{app} \end{pmatrix}. \label{eq:construct_approx}
\end{equation}

The full solution to AMLE may therefore be written as 
\begin{equation} 
\label{eq:solexpansion}
\vu \equiv \vu_{app}^{\e}(t,z;T,Z) + \e \Vreps(t,z)
  \enspace, \end{equation}
where
\begin{equation} 
\Vreps = \begin{pmatrix} 
\Reps_E \\ \Reps_B \\ \Reps_P \\ \Reps_Q
\end{pmatrix} \end{equation}
denotes the error term. To prove the main theorem it suffices to prove that
for any $T_0>0$, $\Vec\Reps$ remains bounded of order one, in an appropriate
norm, uniformly for $\e$ sufficiently small and $0\le t\le T_0/\e$. 

We now derive the equation for $\Vec\Reps$.  Viewing $t,z,T,Z$ as independent
variables,~\eqref{eq:compact}, the equation for $\Vec u^\e$ can be rewritten
as:  
\begin{equation}
\pt \vu = {\cal M}\vu\ - \e{\cal N}\vu +  \w_0^2\efour
 \left(2\e\nu \cos(2k_Bz) P - g (P,z) \right)\enspace ,
\label{eq:mscalecompact}
\end{equation}
where 
\begin{equation}
 {\cal N} =
\begin{pmatrix}
\pT & -\pZ & 0 & 0 \\ -\pZ & \pT & 0 & 0 \\ 
 0 & 0 & \pT & 0 \\
 0 & 0 & 0 & \pT
\end{pmatrix} \enspace.
\end{equation}

To obtain an evolution equation for $\Vreps$, we
substitute~\eqref{eq:solexpansion} into~\eqref{eq:mscalecompact} to obtain
\begin{equation}\begin{split}
\pt (\vu_{app}^{\e}+ \e \Vreps)  = &\ {\cal M}(\vu_{app}^{\e}+ \e \Vreps) \ 
+ \e{\cal N}(\vu_{app}^{\e}+ \e \Vreps) \\
&+  \w_0^2\efour
 \left(2\e\nu \cos(2k_Bz) (P_{app}^{\e}+\e \Reps_P) 
- g (P_{app}^{\e}+\e \Reps_P,z) \right).
\end{split}\end{equation}
We may then eliminate from this two equations obtained during the multiple
scales expansion:
\begin{subequations}
\begin{equation}
\begin{align}
 \bigl( \pt - {\cal M}\bigr) \vu_0 =&\ 0 \enspace;\\
 \bigl( \pt - {\cal M}\bigr) \vu_1 =& -{\cal N}\vu_0 + 
 2\w_0^2\efour\cos(2k_Bz)P_0 + \w_0^2\efour\phi P_0^3
\end{align}\end{equation}\end{subequations}
to leave an equation for the evolution of the error alone
\begin{equation}\begin{split}
\bigl( \pt - {\cal M}\bigr) \Vec R^\e =&\ -\e^{\frac{3}{2}}{\cal N}\vu_1
+ 2\nu\efour\cos(2k_Bz)(\e^{\frac{3}{2}} P_1 + \e R_P^\e)\\
&\ + \w_0^2 \efour \bigl(- \e^{-1}g(P^\e_{app}+ \e R^\e_P,z)
+ \e^{\half}\phi P_0^3
\bigr) \enspace.
\end{split}\end{equation}
To this we add and subtract $\e^{-1} \w_0^2 \efour g(P^\e_{app},z)$ to obtain
\begin{equation}\begin{split}
\bigl( \pt - {\cal M}\bigr) \Vec R^\e =& \ 
-\e^{\frac{3}{2}}{\cal N}\vu_1
+ 2\nu\efour\cos(2k_Bz)(\e^{\frac{3}{2}} P_1 + \e R_P^\e) \\
&+ \e^{-1} \w_0^2 \efour 
\bigl(g(P^\e_{app},z) - g(P^\e_{app}+ \e R^\e_P,z)\bigr) \\
&+\w_0^2 \efour 
\bigl(- \e^{-1}g(P^\e_{app},z)+ \e^{\half} \phi P_0^3 \bigr) \\
=& \  2\e \nu\efour\cos(2k_Bz) R^\e_P\\
&+ \e^{-1} \w_0^2 \efour 
\bigl(g(P^\e_{app},z) - g(P^\e_{app}+ \e R^\e_P,z)\bigr) 
+ \e^{-1} \Vec\rho \enspace ,
\label{eq:R_eqn}
\end{split}\end{equation}
where 
\begin{equation}
\Vec \rho = 
- \e^{\frac{5}{2}}{\cal N} \vu_1 
+ \e^{\frac{5}{2}} \w_0^2 2 \nu \cos{(2k_B z)}P_1 
+ \w_0^2 \left(\e^{\frac{3}{2}} \phi P_0^3 - g(P_{app}^\e,z) \right)
\label{eq:residual}
\end{equation}
is the {\em residual}, essentially the amount by which $u^\e_{app}$ fails to
solve~\eqref{eq:compact}.

We now consider the formal size of the second and third terms on the right
hand side of~\eqref{eq:R_eqn}. Since 
$P^\e_{app}+\e R^\e_P= \e^{\half}\left(P_0 +\e P_1 + \e^{\half} R^\e_P\right)$
and $g(P,z) \sim \phi P^3$, the second 
term of~\eqref{eq:R_eqn} is $\order(\e)$.  We further note that since
$\vu^\e_{app}$ is an approximate solution through order $\e$, the residual
$\vrho$ is formally of order $\e^2$ and so the third term of~\eqref{eq:R_eqn}
is $\order(\e)$. It is in order to control this final term that requires us to
calculate the approximate solution including terms formally of order
$\e^{\frac{3}{2}}$ and also to require that $\Epm$ be in the Sobolev space
$H^3$.
\medskip

\noindent From this discussion, we expect that for times of order $\e^{-1}$,
$\Vreps$ will be bounded. For convenience, we introduce a notation which makes
explicit the expected size of the residual: 
\begin{equation}
\e\vr = \e^{-1}\vrho\enspace.
\end{equation}
 
It is then clear that in order to bound $\Vreps$ we will first need to bound
$\vr$.  

\begin{prop}
({\bf Estimation of the residual})
Let $(\Eplus,\Eminus)$ be a solution of the NLCME system~\eqref{eq:nlcme}.
Then there exists a constant $c>0$ depending on $k$ and the Sobolev
norms of $\Epm$ of order up to three, but independent of $\e$ such that for
all $0\le t \le T_0\e^{-1}$,
\begin{align}
\norm{\vr}{L^\infty} &\le c\  \e^{1/2} \enspace;\\ 
\norm{\vr}{H^1} &\le c \enspace.
\end{align}
\label{prop:r}
\end{prop}

\noindent This proposition is a simple  consequence  of the explicit expression
for $\vr$ (defined in terms of $\vrho$) given in~\eqref{eq:residual}, and of
Prop.~\ref{prop:E}. 

\begin{prop}
Let $(\Eplus,\Eminus)$ be a solution of the NLCME system~\eqref{eq:nlcme}.
Then there exist
$\e_0>0$ and $C_0>0$ s.t. if $0\le \e\le \e_0$, the solution
of~\eqref{eq:R_eqn} satisfies $\norm{\Vreps}{H^1} \le C_0$ for all
$0\le t\le T_0 \e^{-1}$\enspace.
\label{prop:R}
\end{prop}

These propositions imply Theorem~1.

\subsection {Proof of Proposition~\ref{prop:R}: Estimates on the error 
 $\vec R^\e$} 
Recall that the evolution equation for $\Vreps$ is given by:
\begin{equation}
\pt\Vreps = \cM \Vreps  +\efour\w_0^2\left( 2\e\nu \cos(2k_Bz)\Reps_P
-\frac{1}{\e}\left(g (P,z) - g (P^\e_{app},z)\right) \right) + \e \vr
\enspace,
\label{eq:errorequation}
\end{equation}

Motivated by the energy estimates used in Sect.~\ref{sec:ivp} introduce the
weighted norms
\begin{align}
\threeline{\Vreps}^2 
&\equiv \int \Vreps \cdot \Lambda \Vreps \ dz
= \int_{-\infty}^\infty \bigl( {\Reps_E}^2 +{\Reps_B}^2 
+\frac{{\Reps_P}^2}{n^2-1} +\frac{{\Reps_Q}^2}{\w_0^2(n^2-1)} \bigr)
\enspace dz \enspace ; \\ 
\threeline{\Vreps}_{H^1}^2 &\equiv \threeline{\Vreps}^2 + 
 \threeline{\pz\Vreps}^2 \enspace ,
\end{align}
where the weight $\Lambda$ is given by the matrix:
$$
\Lambda = \diag{\left(1,1,\frac{1}{n^2-1},\frac{1}{\w_0^2(n^2-1)}\right)}
$$
These norms are clearly equivalent to the standard $L^2$ and $H^1$ norms. 

Then we have
\begin{equation}\begin{split}
\diff{}{t} \half \threeline{\Vreps}^2 =& 
\frac{1}{n^2-1}
\int \bigl(\ 2 \e \nu \cos(2k_Bz)\Reps_P  +  
\frac{1}{\e} [g(P_{app}^\e,z) - g(P_{app}^\e +\e R_P^\e,z) ]\  \bigr) \Reps_Q
dz   \\ 
 &+ \e \int  \Vreps\cdot\Lambda \Vec{r} \enspace dz.
\end{split}\end{equation}

Similarly, differentiation of~\eqref{eq:errorequation} with respect to $z$ and
left multiplication by $(\pz\Vreps)^{\rm T} \Lambda$ yields:
\begin{equation}\begin{split}
\diff{}{t}\half\threeline{\pz\Vreps}^2 =&
\frac{\e \nu}{n^2-1}
\int \bigl(  2 \cos(2k_Bz)\pz\Reps_P \pz\Reps_Q -
4k_B\sin(2k_B z)\Reps_P\pz\Reps_Q\bigr)\ dz  \\
&+ \frac{1}{\e}\int \bigl( \partial_Pg(P_{app}^\e,z) - 
\partial_P g(P_{app}^\e +\e R_P^\e,z) \bigr)\pz P^\e_{app}\pz\Reps_Q\ dz\\
& - \int \partial_P g(P_{app}+\e\Reps_P,z)  \pz\Reps_P\Reps_Q\ dz\\
&+ \frac{1}{\e}\int \bigl( \pz g(P_{app}^\e,z) - 
\pz g(P_{app}^\e +\e R_P^\e,z) \bigr) \pz \Reps_Q\  dz \\
& + \e \int  \pz\Vreps \cdot \Lambda \pz \Vec{r}\ dz \enspace .
\end{split}\end{equation}

Application of Hypothesis~\ref{hyp:N2}, the $L^\infty$ bound on solutions of
NLCME of Proposition 2, Sobolev's inequality\footnote{$\norm{f}{L^\infty}^2\le
C \norm{f}{L^2}\norm{\pz{f}}{L^2}$}~\cite{J:91} and interpolation yields
\begin{equation}
\label{eq:l2Rbound}
\diff{}{t}\threeline{\Vreps}^2  \le 
C_1 \e \threeline{\Vreps}^2 +
C_2 \e^2\threeline{\Vreps}_{H^1}^4  
+\e \norm{\Vec r}{L^2}\ \threeline{\Vreps}
 \enspace.
\end{equation}
and 
\begin{equation}
\label{eq:dl2Rbound}
\diff{}{t}\threeline{\pz\Vreps}^2  \le C_1 \e \threeline{\Vreps}_{H^1}^2 +
 C_2 \e^2\threeline{\Vreps}_{H^1}^4  + C_3 \e \norm{\pz\Vec r}{L^2}\ 
 \threeline{\Vreps}_{H^1}
 \enspace.
\end{equation}
Estimates~\eqref{eq:l2Rbound} and~\eqref{eq:dl2Rbound} imply
\begin{equation}
\label{eq:h1Rbound}
\diff{}{t}\threeline{\Vreps}_{H^1}^2  \le C\ \bigl( 
 \e \threeline{\Vreps}_{H^1}^2 +
 \e^2\threeline{\Vreps}_{H^1}^4  +\e ||\Vec r||_{H^1}\ 
 \threeline{\Vreps}_{H^1} \bigr)\enspace.    
 \enspace
\end{equation}

If $\Vreps(0)=0$, then by equation~\eqref{eq:errorequation}, 
$\norm{\Vreps}{}\neq 0$ for $t>0$.  We therefore assume 
$\norm{\Vreps(0)}{} >0$. We may then let $\z(t) = \threeline{\Vreps}(t)$.
Then~\eqref{eq:h1bound} and the $H^1$ bound on $\Vec r$ from
Prop.~\ref{prop:r} implies 
\begin{equation}
\diff{\z}{t}  \le C\e \left( 1 + \z + \e\z^3 \right) \enspace. 
 \label{eq:zetaineq}
\end{equation}
This differential inequality is easily solved and we conclude that 
for any $T_0>0$, 
\begin{equation}
||\Vreps(t)||_{H^1} \le  C\ \threeline{\Vreps(t)} \le  C(T_0)
\; \mbox{for }0 \le t \le T_0 \e^{-1} \enspace.
\end{equation}

Finally, we note that, as
\begin{equation}
\vu = \vu_{NLCME} + \e^{\frac{3}{2}} \vu_1 + \e \Vreps \enspace, 
\end{equation}
then
\begin{equation}
\norm{ \vu - \vu_{NLCME}}{H^1} \le \e^\frac{3}{2}\norm{\vu_1}{H^1} +
\e \norm{\Vreps}{H^1}\enspace.
\end{equation}
The estimates on $\Epm$ guarantee that the first term on the right hand side
is $\order(\e)$ and Prop.~\ref{prop:R} guarantees that the second term
is $\order(\e)$.  Thus
\begin{equation}
\norm{ \vu - \vu_{NLCME}}{H^1} \le C \e
\; \mbox{for }0 \le t \le T_0 \e^{-1} \enspace.
\end{equation}
This completes the proof.

\section{Numerical Demonstration--gap soliton propagation and decay}
\label{sec:numerics}

We initialize a wavepacket for AMLE with many oscillations and an envelope
whose form is constructed using the gap soliton solution to the NLCME.  The
gap soliton decays exponentially away from its `center'.  We perform the
simulations with periodic geometry. The period is chosen to be several gap
soliton widths so that the solution is well localized away from the artificial
period ends. The gap soliton initial condition is initialized within the
central region of the domain so the localized structure is essentially
unaffected by the boundary and propagates as though it were on an infinite
spatial domain. In addition, we compute the evolution of a solution to the
NLCME with corresponding envelope initial conditions, and use the
formulae~\eqref{eq:def_E0P0}, \eqref{eq:B0}, and~\eqref{eq:Q0} to construct
approximate solutions to AMLE for comparison.

\subsection{Numerical Methods} 

We use a ``method of lines'' approach, meaning that we first discretize in the
spatial dimension, yielding a set of ordinary differential equations in $t$
for the values of the solution at the discretization points.  We compute
solutions to AMLE as the first order system given in~\eqref{eq:EBPQ}.  We
restrict our computational domain to a finite period interval and discretize
with about 16 points per wavelength. Derivatives are computed spectrally
using discrete Fourier transforms~\cite{GO:77}. That is, suppose ${\cal F}$ is
the discrete Fourier transform, $\xi$ is the dual variable to $z$, and
$\Vec f$ is a vector of discrete values of $f$.  Then the approximate
derivative is given by 
$$
\pz f =\left( {\cal F}^{-1} i \xi {\cal F}\right) \Vec f \enspace.
$$

The spatial discretization of system~\eqref{eq:EBPQ} may now be treated
numerically as a system of ODE's. A fixed-step fourth order explicit
Runge-Kutta method is used to integrate the resulting system in time. Recall
that an $n$-stage explicit Runge-Kutta method for a general evolution equation
$$ \Dot y(t) = f(y,t) $$
is given by~\cite{A:96}:
\begin{align*}
k_1 &= f(y_k, t_k)  \enspace;\\
k_i &= f (y_k + a_i \Delta t, t_k + 
\Delta t\sum_{j=1}^{i-1} b_{i,j} k_j ) \qquad \text{for } i=2,n \enspace ; \\
y_{k+1} &= y_k+ \sum_{i=1}^{n} c_i k_i \enspace.
\end{align*}

Explicit methods tend to impose stability restrictions on the allowable
step size for the time integration.  However, in the case of the
AMLE~\eqref{eq:EBPQ}, this  is simply that $\Delta t < C \Delta z$,
which is a very mild restriction compared, for example, to the heat equation,
for which $\Delta t < C \Delta z^2$.  For most of our simulations we work with
about 20 points per wavelength, which gives $\Delta z \approx \frac{1}{3}$ and
a comparable value for $\Delta t$.

Empirical convergence tests show the method to have fourth-order convergence
in time, and constants of motion are also computed numerically and are shown
to be conserved to 8 or 10 digits.  A similar method is used for NLCME,
though the accuracy is far less crucial as the solutions contain far fewer
oscillations and vary on a slower time scale.

\subsection{Numerical Verification}
To numerically verify and explore the limits of Theorem~1, we solve both AMLE
and NLCME under the SVEA scaling and compare $\vu$ with $\sqrt{\e} \vu_0$ by
monitoring the quantity:
\begin{equation}
{\rm Error}_\e(t) \ =\ \vu^\e\ -\  \sqrt{\e} \vu_0 \enspace.
\end{equation} 
We do this for two values of $\e$, and check that the agreement scales
appropriately as $\e$ is decreased.  This is done for $\e_1=\frac{1}{32}$ and
$\e_2=\frac{1}{64}$, so the error should be reduced by half between the two
runs.   For the purposes of verification, we take much larger values of the
refractive index contrast $\e$ than would be used in a physical experiment. 

\begin{figure} \begin{center}
\includegraphics[width=4in]{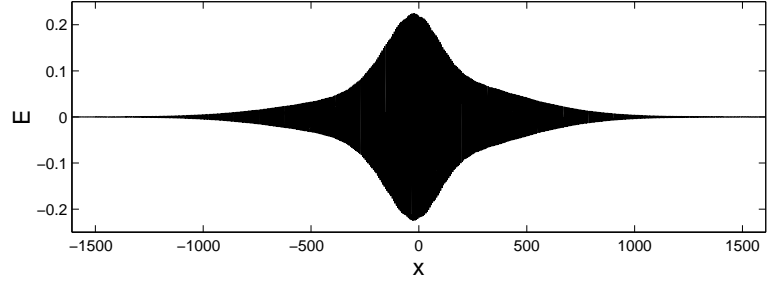}
\caption {The initial data for the electric field satisfying the SVEA.}
\label{fig:efield}
\end{center} \end{figure}

\begin{figure} \begin{center}
\begin{minipage}[b]{.4 \textwidth}
\centering
\includegraphics[width=2in]{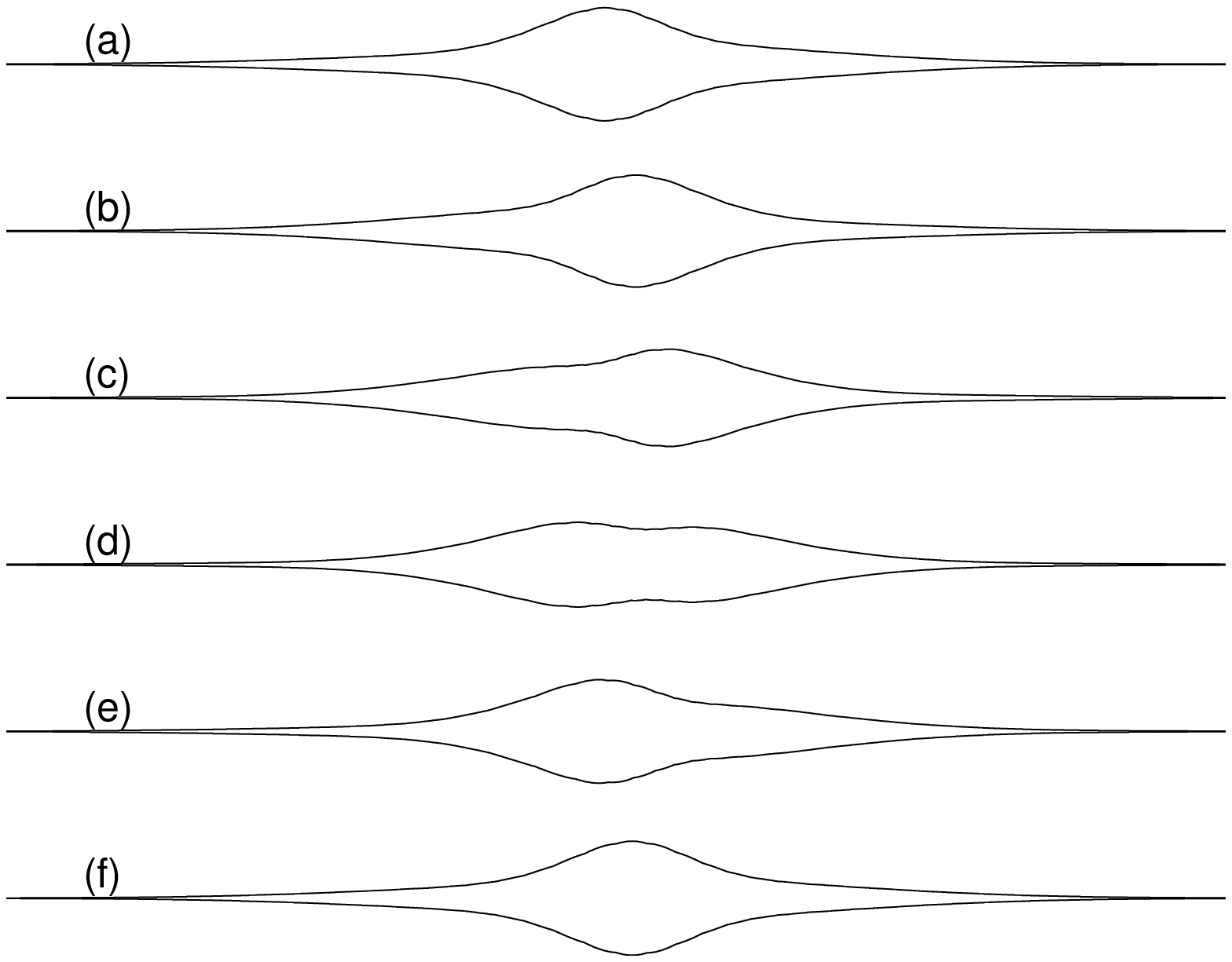}
\par\vspace{0pt}
\end{minipage}
\hspace{.05 \textwidth}
\begin{minipage}[b]{.4 \textwidth}
\centering
\includegraphics[width=2in]{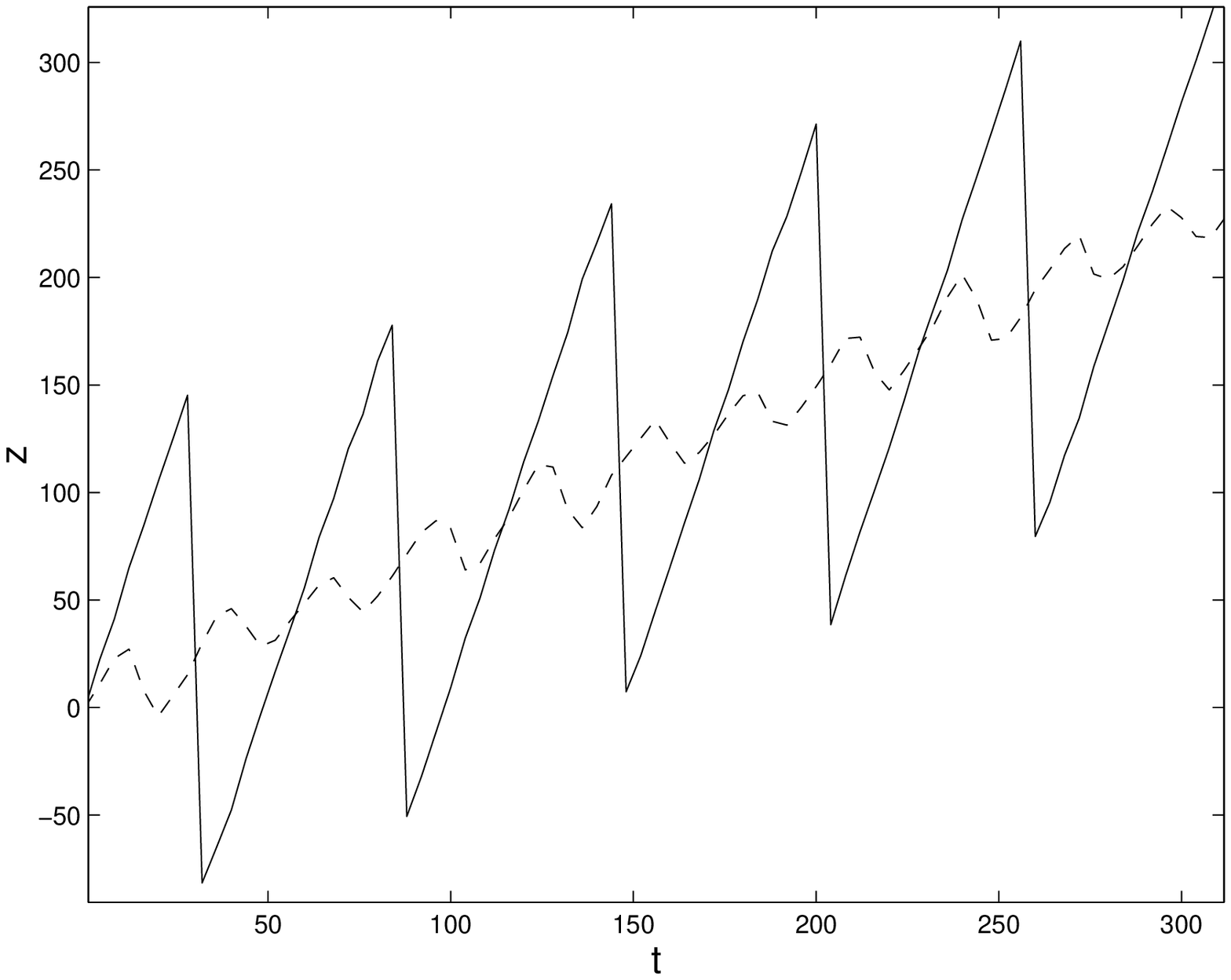}
\par\vspace{0pt}
\end{minipage}
\caption {(Left) The motion of the electric field by reflection and nonlinear
regrouping. Between (a) and (c) the envelope moves forward, between (c) and
(e) it is reflected backwards, and at (f) it has begun propagating forward
again. (Right) The location of the maximum of $E$ (solid line) and the
location of the energy density maximum (dashed line) as a function of time,
showing the effect of reflection off the grating (computed using full AMLE system).}
\label{fig:E_location}
\end{center} 
\end{figure}

In Figure~\ref{fig:efield}, we show a typical initial condition for the
electric field $E$.  The parameters used for this and all the numerical
experiments described in this section are given in
Appendix~\ref{sec:comp_details}. The envelope in this figure is 256
wavelengths long, and it is generated from a simulation with $\e = 1/64$.  The
shape of the electric field envelope as the wave propagates clearly
illustrates the effect of the periodic medium on propagation. In
Figure~\ref{fig:E_location} we see that the electric field envelope (computed
from ``full'' AMLE solutions) is ``two humped'' with the amplitude moving
forward and backward between the humps at a faster rate than the envelope
itself moves forward.  In the same figure, we plot the location of the maximum
of the electric field, and it becomes clear that the electric field maximum
moves forward unsteadily, interrupted by a sequence of backward jumps.  Also
plotted in this figure is the location of the energy density maximum,%
\footnote{The energy density is the integrand on the left hand side of
equation~\eqref{eq:energyestimate}.} 
which propagates more smoothly, since the contribution from the different
fields is averaged.  

\begin{figure} \begin{center}
\includegraphics[width=4in]{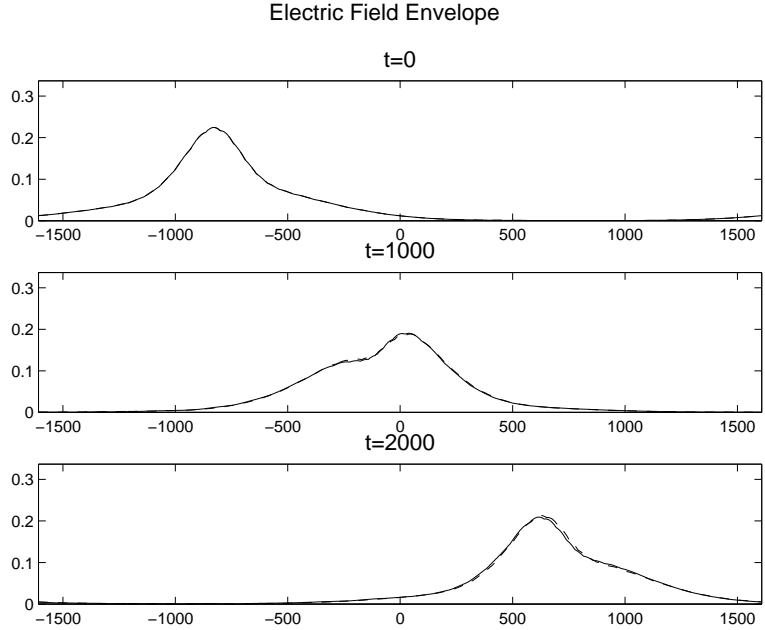}
\end{center} 
\caption {The envelope of the electric field at the beginning, middle, and end
of the computed evolution.  Computations of both the AMLE envelope and its
NLCME approximation are shown.}
\label{fig:E_propagates}
\end{figure}

\begin{figure}\begin{center}
\includegraphics[width=4in]{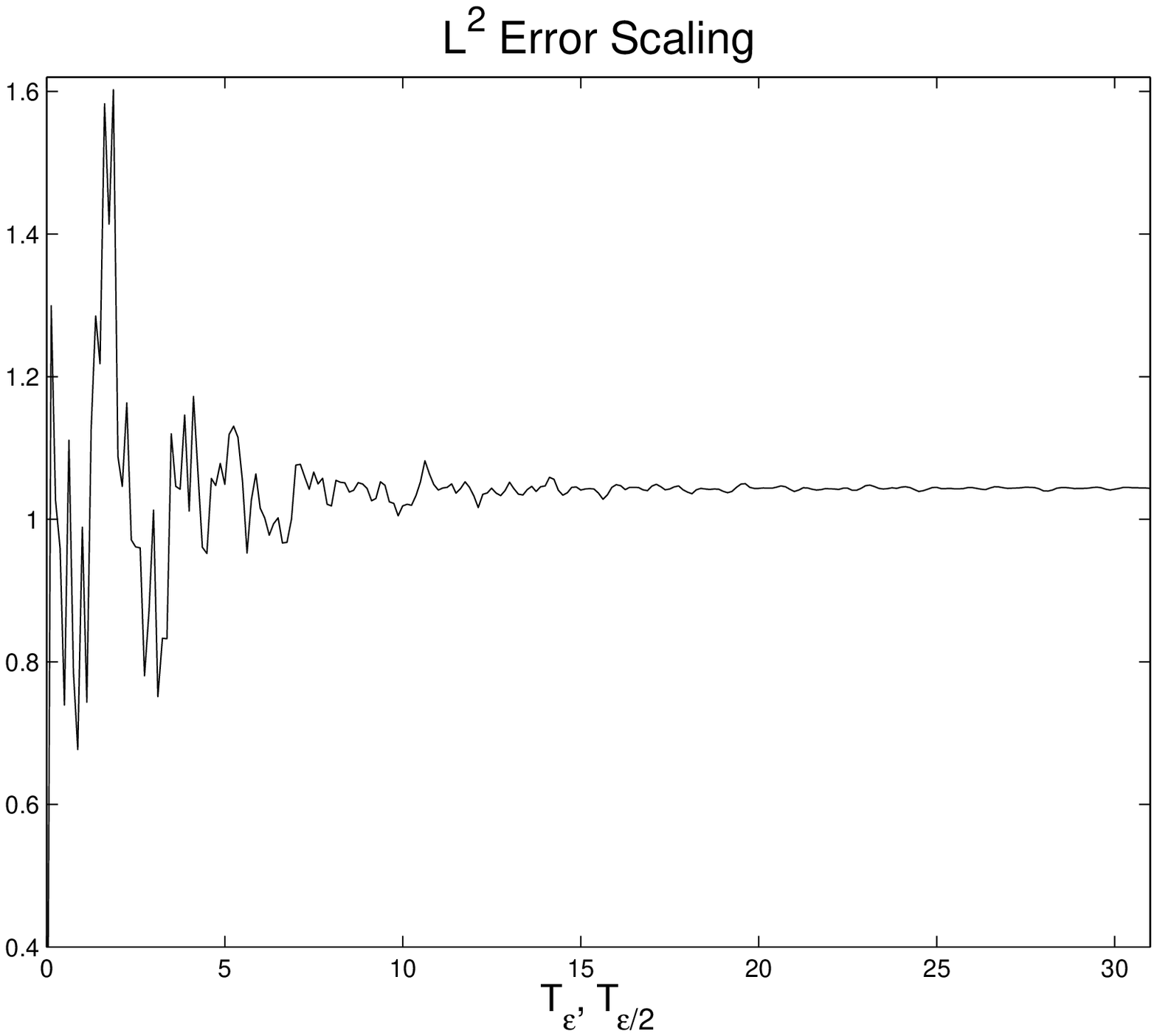}
\caption{Scaling of the error in $L^2$ as a function of the scaled time.}
\label{fig:L2scale}
\end{center}\end{figure}

\begin{figure}\begin{center}
\includegraphics[width=4in]{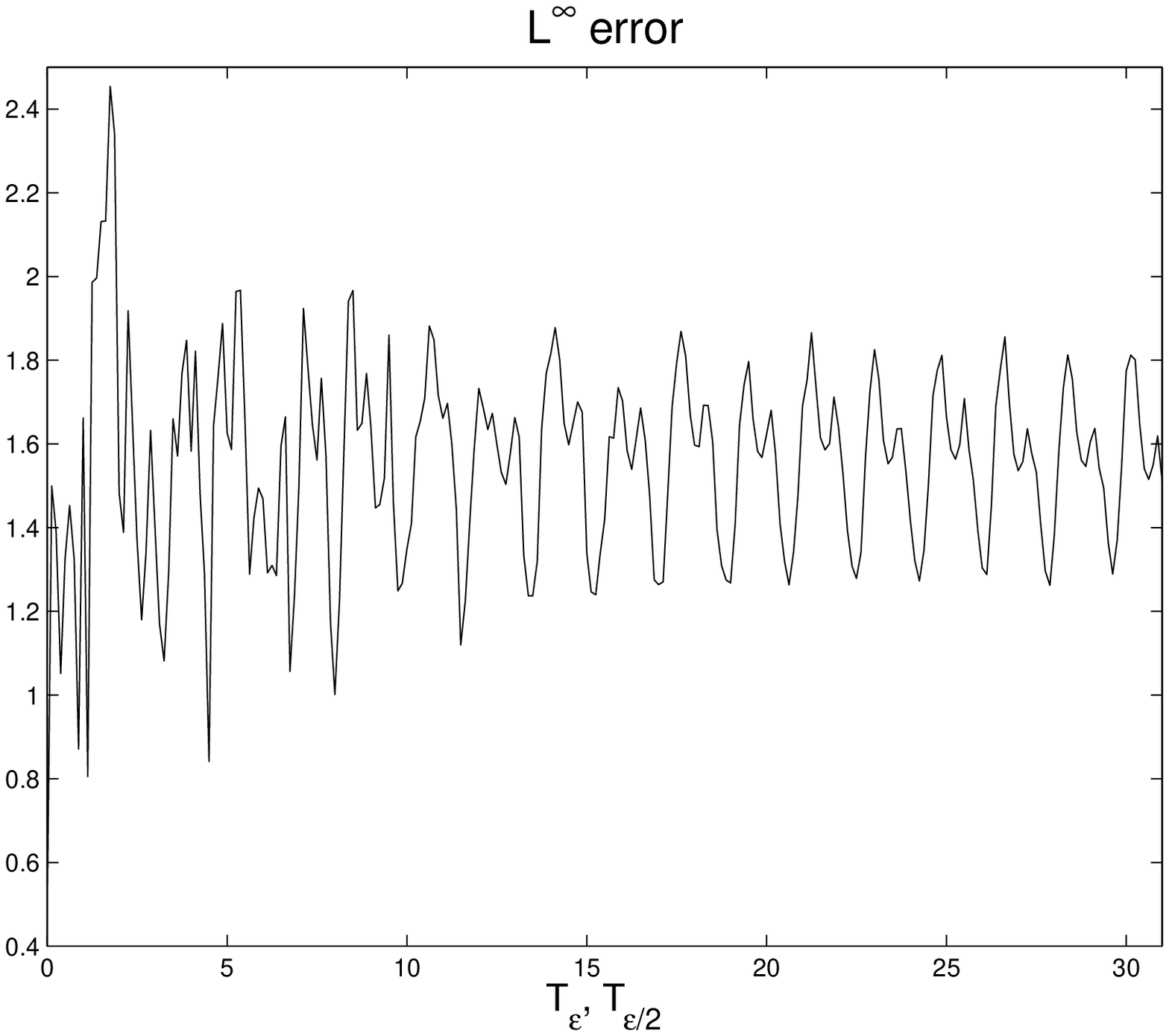}
\caption{Scaling of the error in $L^\infty$ as a function of the scaled time.}
\label{fig:Linfscale}
\end{center}\end{figure}

Figure~\ref{fig:E_propagates} shows the location of the electric field
envelope at the beginning, middle, and end of the computed evolution period.
This figure shows both the envelope computed from the AMLE and also the
approximate envelope computed using the NLCME.  To the ``eyeball metric'', the
agreement appears to be quite close.  More quantitatively, the success of this
procedure is measured by an error-scaling factor given, for any norm,
$$ \mbox{Error Scaling Factor }=
\log_2{ 
\frac{\lVert {Error}_{\e} \rVert(T_\e)}
{\lVert {Error}_{\frac{\e}{2}} \rVert(T_\frac{\e}{2})} 
} 
\enspace.
$$
Then the numerics verify the asymptotic procedure if the scaling factor is
equal to one. Figures~\ref{fig:L2scale} and~\ref{fig:Linfscale} show that
computed in $L^2$ the error scales in agreement with Theorem~\ref{thm:main},
but that the $L^{\infty}$ error is reduced by a factor of $2^{\frac{3}{2}}$.
A general scaling argument shows this is reasonable. Consider a function
$f(z)$ and let $f_\e(z) = \e^{\frac{3}{2}} f(\e z)$, then $\norm{f_\e}{2} = \e
\norm{f}{2}$, while $\norm{f_\e}{\infty} = \e^{\frac{3}{2}} \norm{f}{\infty}$.
It appears that using $H^1$ estimates to control $L^\infty$ estimates has cost
us half a power of $\e$ in our approximation of the error $\Vreps$.
  
\subsection {Very Long Time Behavior}
The error estimates of Theorem~\ref{thm:main} tell us that the solution
constructed from the NLCME and the full solution to AMLE should agree for
times on the order of $\e^{-1}$.  As a practical estimate, this may cause us
some concern, as the width of the solitary wave is also $\order(\e^{-1})$ so
that on these time scales, the distance of propagation is the same order of
magnitude as the width of the solution.  It is therefore of interest to run our
simulations for long times, to see if the NLCME continues to provide a good
approximation beyond what we have proven, or if the approximation breaks down
completely. 

\begin{figure}
\begin{center}
\includegraphics[width=12.2cm]{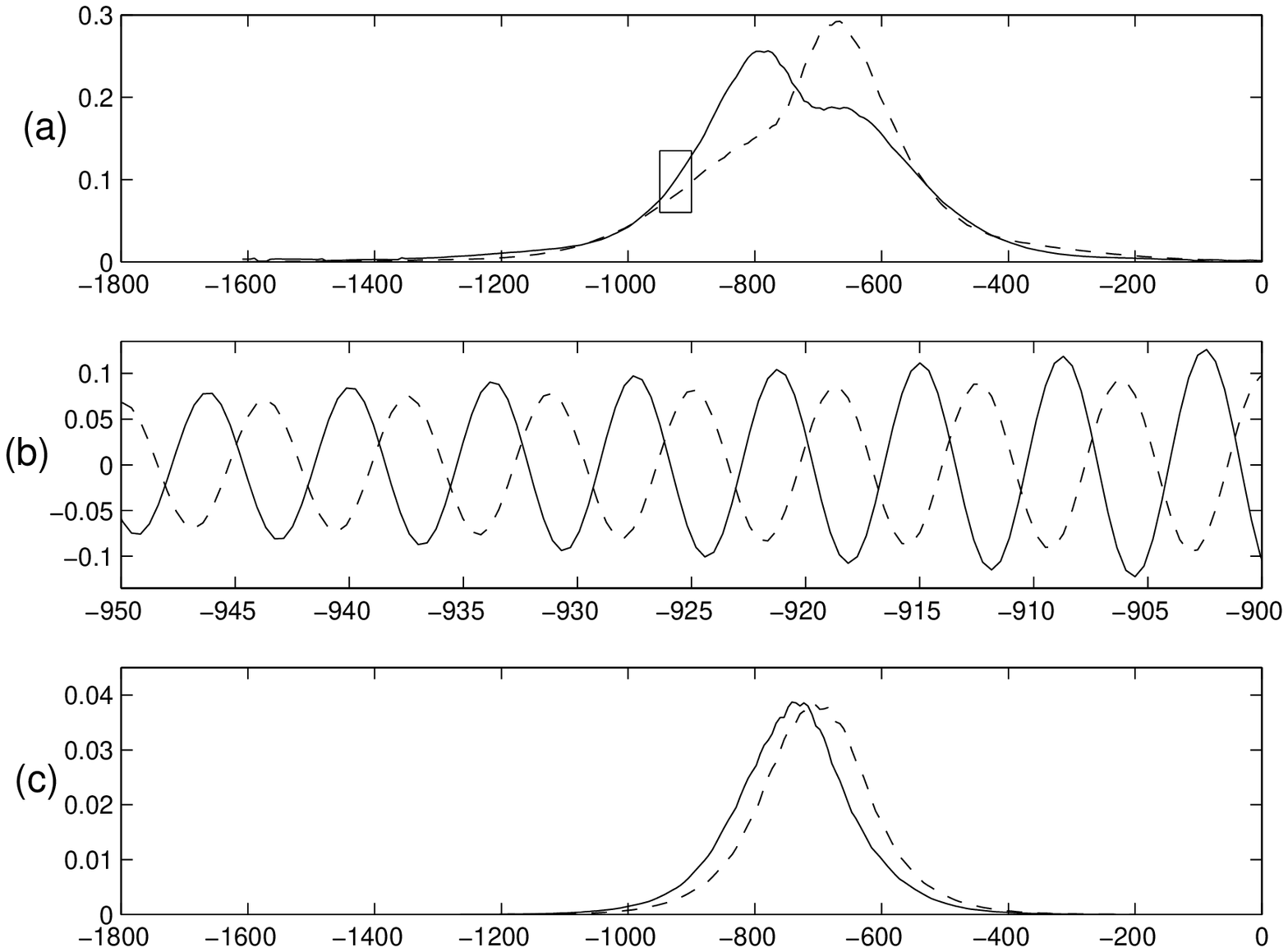}
\caption{ The full (solid) and approximate (dashed) solution at $t=12000$
for {\bf(a)} The electric field envelope, {\bf (b)} the full electric field,
blown up from the box in (a), showing that the two solutions are out of phase,
and {\bf (c)} the energy density which agrees quite well.}
\label{fig:disagreement}
\end{center}
\end{figure}

We run the simulation with $\e= 1/32$, and with $\w_0 = 4$, allowing the
evolution to continue to $t=12000$, which is certainly larger than
$\order(\e^{-1})$. By this time, the $L^2$ norm of the error is of similar to
the $L^2$ norm of the field itself, and has stopped growing.
Figure~\ref{fig:disagreement}(a) shows that the envelopes of the full solution
and the approximation no longer agree, but that they lie in approximately the
same location.  In Figure~\ref{fig:disagreement}(b), we see a blow up of the
electric field and its approximation via the NLCME which shows that the two
solutions are completely out of phase with each other, so that pointwise
estimates will not show any agreement between the solution and the
approximation.  Figure~\ref{fig:disagreement}(c), however, shows that the
energy density of the full solution and the approximation continue to match
very well.  The solution has propagated about twenty times its own width (full
width at half maximum or FWHM), and the centers of the two energy density
plots are separated by about one fifth of a FWHM. This is encouraging, as it
suggests that, although the estimates of Theorem~\ref{thm:main} no longer
hold, the approximation and the full solution have basically remained
together.

\begin{figure}
\begin{center}
\includegraphics[width=12.2cm]{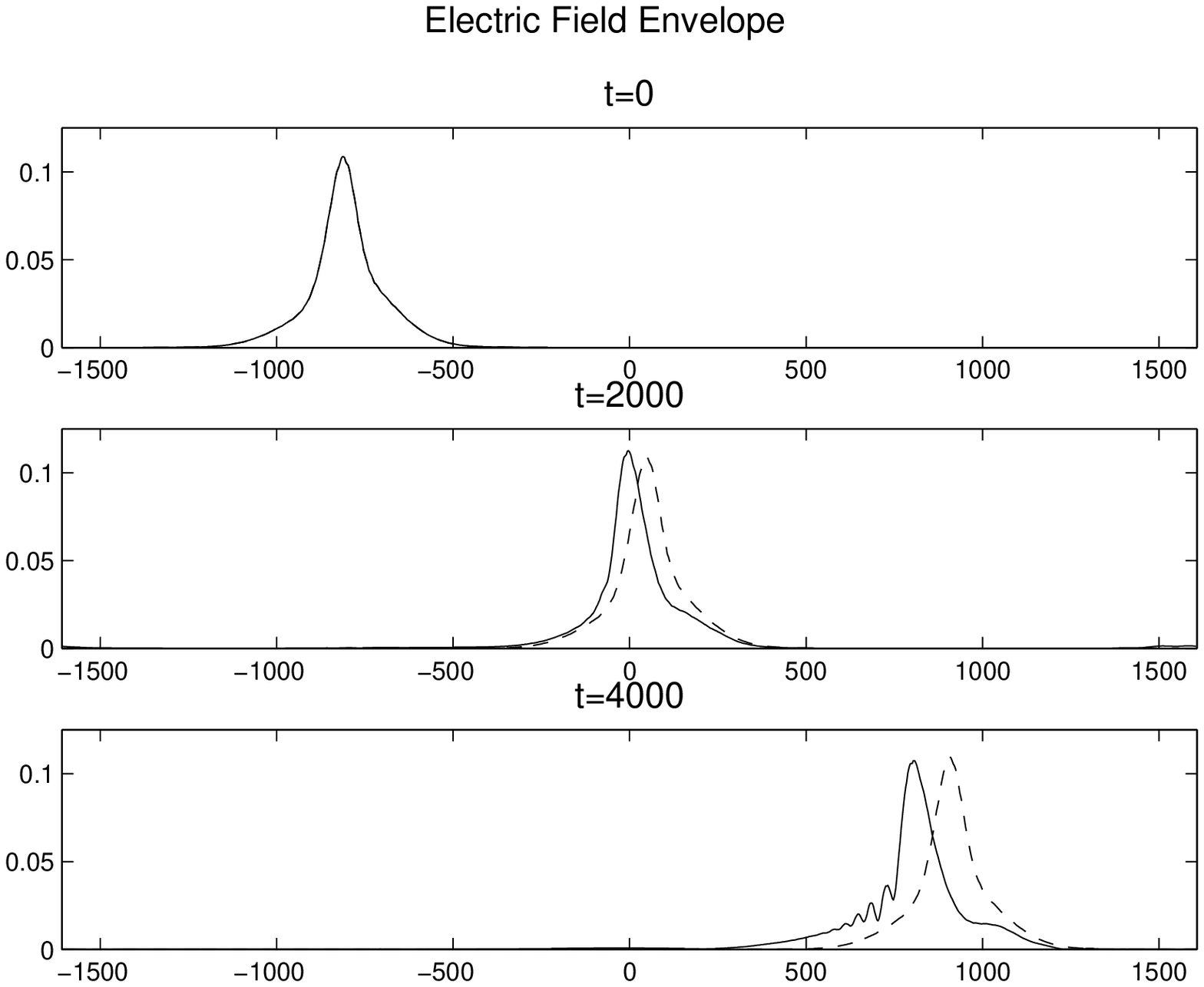}
\caption{The envelope of the solution with very few oscillations steepening
and then breaking up}    
\label{fig:breakup}
\end{center}
\end{figure}

Although the description via the NLCME has broken down in the above
discussion, the electric field has maintained the basic structure of a slowly
modulated plane wave.  Eventually, this very structure will break down and the
solitary wave may itself break apart.  This is shown in
Figure~\ref{fig:breakup} where a solitary wave, moving to the right, steepens
at its trailing end and then begins to break up, while falling behind the AMLE
envelope.  In this figure, the parameters are as in
Figure~\ref{fig:disagreement}, except that $\w_0 =1$, this has the effect of
decreasing the number of oscillations contained within the FWHM of the
envelope from 60 to 10, so that the separation of scales is much less
pronounced. This much narrower envelope breaks up much faster than the
solutions shown in previous plots. Although we have no precise measurement of
this breakup time, it appears to happen on a time scale $t \sim
\order(\e^{-2})$.  

\section{Summary and Discussion}
\label{sec:summary}
In this paper we considered the propagation of high intensity light through a
one-dimensional periodic structure. This was modeled by the anharmonic Maxwell
Lorentz equations (AMLE), which incorporate the effects of material dispersion
due to finite time response of the polarization field, photonic band
dispersion due to the periodic structure, and nonlinearity (intensity
effects). We first gave a detailed discussion of how these effects act
individually and in various combinations, also providing some numerical
illustrations. We next considered AMLE solutions with spatially localized
initial conditions, nearly monochromatic at the Bragg resonant carrier
frequency and such that the effects of dispersion and nonlinearity are
balanced. We proved that, over time scales of interest, the backward and
forward propagating field envelopes satisfy nonlinear coupled mode equations
(NLCME), which we derived from AMLE using multiple scale analysis.
 We also derived rigorous bounds on the deviation of the NLCME solutions
from those of the original Maxwell-Lorentz model. Theorem~1, which describes
this, is the main mathematical result of the paper. We demonstrated its
validity and probed its limitations via numerical simulation, as well as
verifying that the ordering assumptions assumed in our analysis are consistent
with the physical parameter magnitudes characteristic of experimental studies.

Two directions of great interest are the the study of nonlinear phenomena in
multidimensional photonic structures~\cite{AJ:98} and the extension of the
present analysis to the case of more general inhomogeneous structures.  The
multiple scale techniques and the analysis used to obtain Theorem 1 can be
applied to more general structures, e.g. periodic structures defined by a
general Fourier series, index variations which are slow modulations of those
considered, and ``deep gratings''.  In the case of deep gratings, where the
variation of the refractive index is not small, this requires the use of a
multiscale expansion ansatz describing the slow modulation of Floquet-Bloch
waves~\cite{DS:88}, rather than the plane waves we have used in the case of a
system which is nearly translation invariant in $z$.

A number of issues arose in our study we which presently discuss and raise as 
questions meriting further investigation:
\medskip

\noindent {\bf(1)}
Numerical simulations suggest that NLCME continues to acceptably predict the
location of the field energy on time scales for which the estimates used in
proving Theorem~1 break down.  Fig.~\ref{fig:disagreement} shows that the
coupled mode theory fails to predict the location of the individual peaks of
the carrier wave while continuing to predict the location of the energy.  It
would be of interest to investigate whether there exists a weaker, more
general framework, in which the AMLE solution is well described by the NLCME
solution.

\noindent {\bf (2)}
The very long time simulations described in Sect.~\ref{sec:numerics} indicate
a degradation of the gap soliton due to wave steepening and the radiation of
energy away from the soliton core. It would be of interest to derive higher
order model equations which describe these phenomena and agree with the full
solutions to AMLE on longer time scales.

While it is possible to find longer-time envelope equations by starting with
smaller ($\order(\e)$) initial conditions and introducing a third time scale
$T_2= \e^2 t$, our primary interest is to investigate the validity of the
NLCME system already in wide use by experimentalists.  Rigorous results for
such longer-time systems in other contexts are given in~\cite{JMR:98,L:98}.

\noindent {\bf (3)} For the one-dimensional non-dispersive model with
nonlinearity, we have seen that wave steepening and shock formation
occurs. This situation appears to persist in the presence of periodic
structure.  It would be of interest to extend the Lax-Klainerman-Majda
theory~\cite{KM:80} described in Sect.~\ref{sec:aperiodic_nl} to include the
case of equations with periodic or more general inhomogeneous variable
coefficient terms.  As noted, for (localized) SVEA initial conditions with
carrier frequency in Bragg resonance with the medium, photonic band dispersion
is significantly stronger than material dispersion and the gap solitons arise
due to a balance between the former and the Kerr nonlinearity (indeed, the
NLCME may be obtained as a Galerkin truncation of the infinite system of
equations~\eqref{eq:infinite_system}) derived in this case. Inclusion of
material dispersion ($\w_0<\infty$ in AMLE) regularizes shocks; see
Theorem~3. Are there subtle regularizing effects provided by photonic band
dispersion alone?

\noindent {\bf (4)} In the full three-dimensional waveguide problem, one must
also take into account waveguide/mode dispersion and polarization mode
dispersion. In this case there is an interplay between the mechanisms of
diffractive spreading (regularizing), geometric confinement of the field
(tending to one-dimensionalize and therefore singularize the propagation),
modal dispersion (which takes higher harmonics off resonance and therefore
possibly regularizes) and nonlinearity. The interplay of all these effects
remains unclear. It would be interesting to extend the results of~\cite{DR:97,
JMR:98, L:98, S:94a, S:94b} to situations with periodicity and nontrivial
transverse geometry.

\begin{appendix}

\section{Dimensionless Quantities }  
\label{sec:nondim}
In this appendix we nondimensionalize AMLE and isolate the key nondimensional
parameters. We then define {\it dispersion lengths} and {\it nonlinear length}
whose balance specifies the conditions under which a soliton is expected to
form. Finally, using the experimental parameters of Eggleton
et. al.~\cite{E:96}, we calculate our dimensionless quantities and verify the
applicability of AMLE.
\medskip
     
We begin with the AMLE system written using dimensional variables and derive a
nondimensional version of AMLE.  We then use physical parameter values gleaned
from the literature in order to find approximate sizes of the nondimensional
parameters.  Primed variables represent nondimensional quantities and unprimed
variables dimensional ones.  We use the standard notation $[X]$ to represent
the units of $X$ so that $X= [X]X'$ for any variable $X$.

The AMLE written in dimensional variables are:
\begin{subequations}
\begin{align}
&\mu_0 \pt D = \pz B;\;  \pt B  = \pz E \enspace;
\label{eq:dAML1}\\
&D \equiv \eps_0 E+P\enspace ;\label{eq:dAML2}\\
&\tilde\w_0^{-2}\pt^2 P + 
 (1-2 \Delta n \cos(2\tilde k_Bz))P -\tilde\phi P^3 
 = \eps_0 \chi^{(1)}E \enspace.\label{eq:dAML3}
\end{align}
\end{subequations}
 We begin, as usual, by eliminating the magnetic field $B$ to
obtain 
\begin{equation}
\mu_0 \pt^2 D = \pz^2 E \enspace.\label{eq:dAML4}
\end{equation}

We now introduce nondimensional (primed) variables:
\begin{subequations}
\begin{gather}
t = \cT t' ; \;
z = \cZ z' \enspace;\\
E = \cE E' \enspace; \;
P = \eps_0 \cE P' \enspace; \;
D = \eps_0 \cE D'\enspace; \label{eq:EPDscale} \\
\tilde k_B = \frac{k_B}{\cZ} \enspace; \;
\tilde \w_0 = \frac{\w_0}{\cT} \enspace; \;
\label{eq:kwphiscale}
\end{gather}
\end{subequations}
where the calligraphic letters represent dimensional magnitudes.  
 To explicitly display the expected scaling, we write
\begin{equation}
  \Delta n = \varepsilon \nu \ {\rm{and}} \ 
\tilde \phi = \frac{\e \phi}{(\eps_0 \cE)^2} \enspace,
\label{eq:dnphiscaling}
\end{equation}
where $\nu$ and $\phi$, along with $\chi^{(1)}$, are dimensionless
and ${\cal O }(1)$, and the fields in (\ref{eq:EPDscale}) are also all
${\cal{O}}(1)$.

Substituting these new variables into
equations~\eqref{eq:dAML4},~\eqref{eq:dAML2}, and~\eqref{eq:dAML3} and
eliminating common factors yields:
\begin{subequations}
\label{eq:AMLE_dimless}
\begin{align}
&\frac{\mu_0 \eps_0}{\cT^2} \partial_{t'}^2{D'}
= \frac{1}{\cZ^2}\partial_{z'}^2{E'}\enspace ; \label{eq:D_evolve} \\
&D' = E' + P'\enspace ; \label{eq:Ddef} \\
&\frac{1}{{\w_0}^2} \partial_{t'}^2 P' + [ 1 + 2 \e \nu\cos(2 k_B z')] P' - 
\e \phi {P'}^3 = \chi^{(1)} E' \enspace.
\label{eq:P_eqn_nondim}
\end{align}
\end{subequations}
Letting  
\begin{equation}
\cT = \frac{\cZ}{c} \enspace,
\end{equation}
we have that~\eqref{eq:D_evolve} becomes 
\begin{equation}
\partial_{t'}^2 D' = \partial_{z'}^2 E'\enspace. \label{eq:wave_eq_nondim}
\end{equation}
The system~\eqref{eq:wave_eq_nondim}, \eqref{eq:Ddef},
and~\eqref{eq:P_eqn_nondim} comprise the dimensionless AMLE system; see 
also~\eqref{eq:AMLE}.  

\subsection{ The material frequency ${\tilde\w}_0$ and the electric
susceptibility $\chi^{(1)}$}

At low intensities the relation between $P$ and $E$ is given by the 
{\it Lorentz} model:
\begin{equation}
\frac{1}{\tilde\w_0^2} P_{tt} + P =  \eps_0 \chi^{(1)} E\enspace.
\end{equation}
 In the time-frequency domain this implies:
\begin{equation}
\Hat{P}(\w) = \epsilon_0\frac{\tilde\w_0^2}{\tilde\w_0^2-\w^2}\ \Hat{E}(\w)
\enspace, 
\label{eq:lorentzchi}
\end{equation}
where $\Hat{f}(\w)=\int e^{-i\w t}f(t)\ dt$. 

In a general linear setting we have, 
\begin{equation}
\Hat{P}(\w) = \chi^{(1)}(\w ) \Hat{E} \enspace ,
\end{equation}
where the (frequency dependent) 
index of refraction, $n(\w )$,  is related to $\chi^{(1)}$ by the relation:
\begin{equation}
 n^2(\w )\ =\ 1 + \chi^{(1)}(\w ) \enspace .
\label{eq:nofomega}
\end{equation}
A standard model for  $\chi^{(1)}(\w )$ in the optics
literature~\cite{A:95} is the Sellmeier model, which approximates
$\chi^{(1)}(\w )$ by a function of the form: 
\begin{equation}
\chi^{(1)}(\w) = \eps_0 \sum_{i=1}^N 
\frac{ \w_i^2 \chi^{(1)}_i}{\w_i^2 - \w^2} \enspace,
\label{eq:sellmeier}
\end{equation}
where $\w_i$, the model resonant frequencies of the medium and $\chi^{(1)}_i$
are determined by a data fit.  For silica glass, a good fit with experimental
data is found with $N=3$. The Lorentz model corresponds to $N=1$, so we take
the term in the $N=3$ expansion corresponding to that frequency, $\w_i$ which
is closest to the input carrier frequency.  Below, we use this to determine
the values of $\tilde\w_0$ and $\chi^{(1)}$ in the Lorentz model.

\subsection{ The Electric Field Strength $\cE$}
Most optical physics literature reports field strength in terms of the
intensity, $I$. The electric field strength is given in terms of the intensity
by~\cite{A:95}:
\begin{equation}
\cE^2 = \frac{2 I}{\eps_0 c n} \enspace.
\end{equation}
where $n$ is the (nondimensional) refractive index, related to the linear
susceptibility, $\chi^{(1)}$ by~\eqref{eq:nofomega}.

\subsection{The coefficient of nonlinearity, $\tilde\phi$.}
We consider the instantaneous limit of the basic equation, with no grating,
i.e. $\Delta n=0$:
\begin{equation}
P - \tilde\phi P^3 = \eps_0 \chi^{(1)} E\enspace.
\end{equation}
We may invert the above relation for small $E$ and write 
\begin{equation}
P = \eps_0 \left(\chi^{(1)} E + \chi^{(3)} E^3 + \dots \right) \enspace ,
\end{equation}
where 
\begin{equation}
\tilde\phi = \frac{\chi^{(3)}}{\eps_0^2 {\chi^{(1)}}^3}\enspace.
\end{equation}
Then the nondimensional quantity is given by
\begin{equation}
 \e \phi = \frac{\cE^2\chi^{(3)}}{ (\chi^{(1)})^3}\enspace.
\end{equation}
The third order susceptibility $\chi^{(3)}$ is related to the nonlinear
refractive index, $n_2$ or $n_2^I$ by the relation (\cite{A:95}, page 40,  
equation (2.3.13) and page 582, equation (B.2)):
\begin{equation}
 \chi^{(3)} = \frac{8 n n_2}{3} 
= \frac{4 \epsilon_0 c n^2 n_2^I}{3}\enspace.
\end{equation}
Finally, since $I=\half\epsilon_0cn\cE^2$ (\cite{A:95}, page 582,
equation (B.1)), we have
\begin{equation}
\label{eq:phiparam}
 \e \phi =\frac{8 I n n_2^I }{3(\chi^{(1)})^3}
\enspace.
\end{equation}

\subsection{Parameter values of physical experiments}
To form the anharmonic oscillator equation for the polarization, we need four
constants: the susceptibility $\chi^{(1)}$, the nondimensional frequency
$\w_0$, the index modulation $\Delta n$, and the cubic coefficient $\phi$.

\noindent{\bf The Susceptibility $\chi^{(1)}$ and the nondimensional frequency
$\w_0$}\\  
\noindent First we must find the characteristic time scale ${\cal T}$. Typical
experiments are performed using laser light with wavelength of approximately
one micron.  We define the characteristic length and time so that $k_B\approx
1$, but for convenience in the paper refer to $k_B$.  Accordingly, we take
\begin{alignat}{2}
\cZ &= \frac{1 \times 10^{-6}}{2\pi}\mbox{m }   &\approx 1.6 \times 10^{-7}
\mbox{m}\enspace ;\\
\cT &= \frac{\cZ}{c} &\approx 5.3 \times 10^{-16} \mbox{s}\enspace.
\end{alignat}

Next, we must find the dimensional frequency of the oscillator. 
 For silicon glass, one has (\cite{A:95}, page 7) 
\begin{align}
\tilde\w_0 &= 1.6 \times 10^{16} \mbox{s}^{-1} \enspace; \\
\chi^{(1)} &= .41 \enspace. \label{eq:chi_approx}
\end{align}
The nondimensional resonant frequency is then given by
\begin{equation}
\w_0 = {\tilde \w}_0 {\cal T} \approx 8.6\enspace.
\end{equation}

\noindent{\bf The index modulation $\Delta n= \e\nu$}\\
\noindent Eggleton et. al.~\cite{E:96} give an approximate value of
\begin{equation} \Delta n \approx 3 \times
10^{-4}\enspace. \label{eq:def_delta_n}  
\end{equation}

\noindent{\bf The nondimensional nonlinearity coefficient, $\phi$}
\noindent 
For this we need the intensity, which in~\cite{E:97} is given by
\begin{equation}
I \approx 2\times 10^{14} \ \mbox{W}/{\mbox{m}^2} 
\end{equation}
and the nonlinear refractive index (\cite{A:95}, pages 582--583):
\begin{equation}
n_2^I = 2.5 \times 10^{-20} \ \mbox{m}^2/\mbox{W} \enspace.
\end{equation}
The linear refractive index is obtained from~\eqref{eq:nofomega}
and~\eqref{eq:chi_approx}: 
\begin{equation}
n \approx 1.2 \enspace.
\end{equation}
From~\eqref{eq:phiparam} we have 
\begin{equation}
 \e \phi\approx 2\times 10^{-4} \enspace.
\end{equation}
Therefore, by choosing the small parameter
\begin{equation}
\e = 10^{-4} \enspace,
\end{equation}
we arrive at
\begin{equation}
\phi \approx 2
\end{equation}
and
\begin{equation}
\nu \approx 3 \enspace .
\end{equation}
Therefore the  approximate nondimensional polarization equation 
\eqref{eq:P_eqn_nondim} may be
written: 
\begin{equation}
\order(10^{-2}) \partial_{t'}^2 P' + 
 [\ 1 + \order(10^{-4}) \cos (2 k_B z')\ ]P' 
 + \order(10^{-4}) {P'}^3 = \order(1) E' \enspace.
\end{equation}
We see that the nonlinearity and the dimensionless grating effectively balance
each other.  This justifies our $\e$-dependent scaling of the dimensionless
AMLE system and the solution.  We note that this scaling assumes that $E'$,
$P'$, and $D'$ are $\order(1)$ quantities, see~\eqref{eq:EPDscale}.  In the
main text, we take $E$, $P$, and $D$ to be $\order(\sqrt{\e})$, thereby
effectively introducing the factor of $\e$ multiplying $\phi$
in~\eqref{eq:P_eqn_nondim} which is absent from~\eqref{eq:AMLE2}.

Note also, that while ${\w_0}^{-2}$, the coefficient of $\partial_{t'}^2 P'$,
is small, it is roughly 100 times the grating strength, {\it i.e.}
$\e\w_0^2\sim 10^{-2}$.  The significance of this can be seen as follows.
Were we to expand the electric field as in Sect.~\ref{sec:derivation}, to all
orders in $\e$ we would have:
$$ E \approx \sqrt{\e} \sum_{i=0}^{\infty} \e^{i}E_i \enspace. $$
Inspection of the hierarchy of equations for $E_i$ reveals that 
$$ E_i \sim {\w_0}^{2i}. $$ This suggests that $E^\e$ is well-approximated by
$\e^\half E_0$ provided $\e {\w_0}^2 \ll 1$. The experimental regime discussed
satisfies this criterion.

\section{Calculation of the dispersion and nonlinear lengths}

In the design of an experiment to observe gap solitons, the matter of the  
{\it formation length} is important. Laser light injected into an optical
fiber will have an approximately Gaussian profile. One is therefore interested
in the distance over which one can expect a soliton to form. Solitons are
understood to form due to a balance of dispersive and nonlinear effects. 
Dispersion acts by broadening a pulse and radiating  high frequency components
away, while a Kerr (focusing) nonlinearity acts to concentrate energy. We
presently give a heuristic discussion of this balance.

\subsection*{Material dispersion length, $z_{\rm D,material}$} 
Recall that the (material) dispersion relation associated with the finite time 
response of the medium to the field is:
\begin{equation}
k^2\ =\ \w^2 \frac{n^2 - \bigl(\frac{\w}{\w_0}\bigr)^2}
                  {1- \bigl(\frac{\w}{\w_0}\bigr)^2} \enspace .
\end{equation}
The dispersion of a wave packet, with frequency content concentrated in 
 an interval of width $\e$ about $\w_B$ is governed by Fourier integrals of 
the form:
\begin{equation}
I(z,t)\ =\ \int e^{i(k(\w)z'-\w t')}\ f(\frac{\w-\w_B}{\e})\ d\w \enspace,
\end{equation}
where $f$ is a localized function of frequency.
Expansion of $k(\w)$ about $\w=\w_B$  yields
\begin{equation}\begin{split}
I(z',t') &\sim e^{i(k_Bz'-\w_Bt')} \int e^{i(\w-\w_0)(k'(\w_B)z'-\w_Bt')}
e^{i\frac{k''(\w_B)}{2}(\w-\w_B)^2t'}\ f(\frac{\w-\w_B}{\e})\ d\w\\
&\sim \e
e^{i(k_Bz'-\w_Bt')} \int e^{i\mu(k'(\w_B)(\e z')-\w_B (\e t') )} 
e^{i\frac{k''(\w_B)}{2}\mu^2(\e^2z')}\ f(\mu)\ d\mu \enspace\\
&= {\cal O}( (k''(\w_B)\e^2z')^{-\half} )\enspace ,\ \ \ 
z'= {\cal O}(\e^{-2})\enspace .
\end{split}\end{equation}
Thus, a localized pulse disperses due to the finite time response of the
medium over a dimensionless  distance $z'$ of order $\e^{-2} k''(\w_B)^{-1}$. 
Noting that $k''(\w_B) = {\cal O}(\w_0^{-2})$, we have:  
$$ 
z_{\rm D,material}= \order(\e^{-2} k''(\w_B)^{-1}) = 
 \order(\w_0^2\e^{-2}) \text{ wavelengths}.
$$
Using the physical parameter values discussed in Appendix~\ref{sec:nondim}, we
find that
$$ z_{\rm D,material} \approx 7 \ \rm{km} \enspace . $$ 

\subsection*{Photonic band dispersion length, $z_{\rm D,band}$}

Linear dispersion due to the periodic structure (photonic band dispersion) is
governed by the linear coupled mode equations~\eqref{eq:linearcme}.  The
dispersion of the wave envelope is then expressed in terms of generalized
Fourier superpositions of Floquet-Bloch waves:
\begin{equation}
\begin{split}
I(z',t') & \sim \int E(z',t',;k_B+\e Q) f(Q) dQ \\
& \sim    e^{i(k_Bz'-\w_Bt')}\ e^{-i\Omega(0)\e t'}\
  \int e^{iQ(\e z' - \Omega'(0)\e t')}\ e^{-{\frac{i}{2}}\Omega''(0)Q^2\e t'}
  f(Q) dQ \\
&= \order( (\Omega''(0)\e t)^{-\half}  )\enspace ,\ \ \ 
t'=\order( \e^{-1} ) \enspace,
\end{split}
\end{equation}
where $\Omega(Q)$ is given by the dispersion relation~\eqref{eq:OmegaQ}  
which disperses to zero over a distance 
$$
z_{\rm D,band} =
\order([\e\Omega''(0)]^{-1})\ = \order(\frac{\k}{\e}) \text{ wavelengths}.
$$
Physically, this gives 
$$
z_{\rm D,band} \approx 1 \ { \rm cm}
$$
which is six orders of magnitude shorter than $z_{\rm D,material}$.

\subsection*{Nonlinear length, $z_{\rm NL}$} 

A measure of the distance, $z_{\rm NL}$, over which nonlinear effects play a
role can be obtained by considering the coupled mode equations in the absence
of dispersion. If $E_0$ denotes the electric field amplitude then we have:
\begin{equation} (\partial_T\pm v_g\partial_Z)E = -i\G |E|^2 E
\end{equation}
with solution
$E = e^{-i\G \cE^2(Z-v_gT)}\cE= e^{-i\G \cE^2(\e z'-v_g\e t')}\cE$, 
for some constant $\cE$.
Therefore,  
$$
z_{\rm NL} = \left( \e\G\right)^{-1} 
 \sim \left(\e \phi \right)^{-1} \text{ wavelengths}
$$ 
which gives 
$$
z_{\rm NL} \approx 1 \ {\rm cm}
$$ 
which balances the band dispersion length $z_{\rm D,band}$.

\subsection*{Balance of nonlinearity and dispersion}

Note that $z_{\rm D,material}$ is longer than $z_{\rm D,band}$ by a factor of
order $\e^{-1}$; for frequencies near the band edge, the dispersion due
to the periodic structure is much stronger than material dispersion.

Therefore, in order to achieve a balance between dispersive and nonlinear
effects over a short distance we must equate $z_{\rm D,band}$ and 
$z_{\rm NL}$.  This gives
\begin{equation}
\frac{\k}{\e} \sim  \frac{1}{\phi} \text{ or } \k\phi\sim \e \enspace.
\end{equation}
By~\eqref{eq:phiparam} this is gives the intensity ensuring a balance of 
appearance of nonlinear effects within a (photonic band) dispersion length,
$z_{\rm D,band}$:
$$
I\ \sim\ \frac{3}{8} \frac{(\chi^{(1)})^3 \e}{n n_2^I\k} \enspace
$$
which works out to 
$$
I = \order(10^{14})\ \mbox{W}/{\mbox{m}^2} \enspace,
$$
in line with the experiments described in~\cite{E:97}.

\section{Computational Details}
\label{sec:comp_details}
\subsection{Computations in Sects.~\ref{sec:aperiodic_nl} 
and~\ref{sec:nl_finite_time}} All figures in these sections are initialized as
a single normal mode of wavenumber $k=1$ of the linearized form the
AMLE,~\ref{eq:AMLE}, with $\nu=0$ and magnitude $\sqrt{\e}$. We should note
that the numerical simulations use very large values of $\e$ compared to the
physically appropriate value $\e = \order(10^{-4})$ derived in
Appendix~\ref{sec:nondim} because performing the simulations for AMLE with
such small $\e$ would be computationally infeasible.  The other parameters are
given by
$$ n^2-1 =1 \text{ and } \phi =1\enspace.$$ 
For Fig.~\ref{fig:shock}, we use a frequency of $\w_0= 1000$ to illustrate the
behavior near the limit of instantaneous polarization before the onset of a
shock.  In Figs.~\ref{fig:small_omega} and~\ref{fig:big_omega}, we use
$\w_0=50$ and $\w_0=100$, respectively to show the role of dispersion in
regularizing the shock.

\subsection{Computations in Sect.~\ref{sec:numerics} }
All calculations in this sections are performed with the following parameter
values:
\begin{align*}
\e &= \frac{1}{32} \enspace;\\
k_B &= 1 \enspace ;\\
n &= 1.19 \enspace ; \\
\phi &=1 \enspace ;\\
\nu &=1 \enspace.
\end{align*}
In Figs.~\ref{fig:efield}, \ref{fig:E_propagates}, and~\ref{fig:disagreement},
a material frequency of 
$$\w_0=4$$ is used, while in Fig.~\ref{fig:breakup}, we use the value
$$\w_0=1 \enspace.$$
In all calculations, the coefficients of the NLCME are derived from the above
parameters, and, as initial conditions, we construct a solution from the gap
soliton with parameters
$$ v = .9 \text{ and } \d = .9 \enspace. $$

To create the graphs in Figs.~\ref{fig:L2scale} and~\ref{fig:Linfscale}, we
also compute the evolutions with all parameters as above except with
$\e = \frac{1}{64}$.

\end{appendix}

\section*{Acknowledgements}

\noindent R.H. Goodman was supported by an NSF University-Industry
postdoctoral fellowship DMS-99-01897.  P.J. Holmes was partially supported by
DOE grant DE-FG02-95ER25238. The authors wish to acknowledge informative and
stimulating conversations with the following individuals: Alejandro Aceves,
Ben Eggleton, Dick Slusher, Gadi Lenz, Mel Lax, Eduard Kirr, Stefan Spalter,
Peter Oswald, Steve Golowich and an anonymous referee for pointing out
numerous relevant references.

\nocite{X:00}
\bibliographystyle{amsplain}
\bibliography{gwh-revised}
\end{document}